\documentclass[aps, prd, showpacs, floatfix, superscriptaddress, twocolumn, nofootinbib, preprintnumbers, longbibliography]{revtex4-2}
\usepackage{lipsum, multirow, microtype, amsmath, amssymb, newfloat, bm, color}
\usepackage{graphicx}
\usepackage{natbib}
\usepackage{comment}
\usepackage{hyperref}
\usepackage{hypcap, mathrsfs, placeins, etoolbox}
\usepackage{dcolumn}
\usepackage[dvipsnames]{xcolor}
\usepackage[utf8]{inputenc}
\usepackage[scaled = 0.92]{helvet}
\usepackage{inconsolata}
\usepackage{enumitem}
\usepackage{orcidlink}
\usepackage{xr}
\usepackage{cleveref}
\setlist[itemize]{leftmargin = 9pt, labelsep = 0.31em, itemsep = 0.055em}
\graphicspath{{Figures/}} 
\hypersetup{
colorlinks = true,
citecolor = cyan,
linkcolor = magenta,
urlcolor = NavyBlue,
allbordercolors = {0 0 0},
pdfborderstyle = {/S/U/W 1}
}

\def \GW        {\text{GW}}
\def \LIGO      {\text{LIGO}}

\def \SNR       {\text{SNR}}

\def \BBH       {\text{BBH}}

\def \O         {\text{O}}

\def \CNN {\text{CNN}}

\def \PSD {\text{PSD}}
\def \AS {\text{AS}}
\def \PS {\text{PS}}

\def \GW {\text{GW}}
\def \IMRPhenomD    {\text{IMRPhenomD}}
\def \IMRPhenomPv    {\text{IMRPhenomPv}2}
\newcommand{\IAR}{Institute of Advanced Research, Koba Institutional Area, Gandhinagar - 382 426, India.\vspace*{0.125cm}}
\newcommand{\SXC}{St.Xavier's College (Autonomous), Navrangpura, Ahmedabad - 380 009, India.\vspace*{0.125cm}}
\newcommand{\KAAS}{Kshama Ahmedabad Academy of Sciences, Motera, Ahmedabad - 380 005, India. \vspace*{0.125cm}}

\newcommand{\QWT}{Quantum Walks Technologies, Morris Plains, NJ 07950, USA.\vspace*{0.125cm}}
\newcommand{\NIKHEF}{Nikhef, Science Park 105, 1098 XG Amsterdam, The Netherlands.}

\newcommand{\IWF}{Space Research Institute, Austrian Academy of Sciences, Schmiedlstrasse 6, 8042 Graz, Austria}
\newcommand{\UMD}{Department of Physics, University of Massachusetts, Dartmouth, MA 02747, USA.}

\begin{document}
\title{Detection of Gravitational Wave Signals from Precessing Binary Black Hole Systems using Convolutional Neural Network
}
\author{Chetan Verma} 
\email{chetanverma.phd@iar.ac.in} \affiliation{\IAR}
\author{Amit Reza {\orcidlink{0000-0001-7934-0259}}} 
\email{amit.reza@oeaw.ac.at} \affiliation{\IWF}
\affiliation{\NIKHEF} 
\author{Gurudatt Gaur \vspace*{0.15cm}} 
\email{gurudatt.gaur@sxca.edu.in} \affiliation{\SXC} 
\affiliation{\KAAS}
\author{Dilip Krishnaswamy \vspace*{0.15cm}} 
\email{dilip@ieee.org} \affiliation{\QWT}
\author{Sarah Caudill \vspace*{0.15cm}} 
\email{scaudill@umassd.edu}
\affiliation{\UMD}
\begin{abstract}
Current searches for gravitational waves (GWs) from black hole binaries using the LIGO and Virgo observatories are limited to analytical models for systems with black hole spins aligned (or anti-aligned) with the orbital angular momentum of the binary. Detecting black hole binaries with precessing spins (spins not aligned or anti-aligned with the orbital angular momentum) is crucial for gaining unique astrophysical insights into the formation of these sources. Therefore, it is essential to develop a search strategy capable of identifying compact binaries with precessing spins. Aligned-spin waveform models are inadequate for detecting compact binaries with high precessing spins. While several efforts have been made to construct template banks for detecting precessing binaries using matched filtering, this approach requires many templates to cover the entire search parameter space, significantly increasing the computational cost.
This work explores the detection of GW signals from binary black holes (BBH) with both aligned and precessing spins using a convolutional neural network (CNN). We frame the detection of GW signals from aligned or precessing BBH systems as a hierarchical binary classification problem. The first CNN model classifies strain data as either pure noise or noisy signals (GWs from BBH). A second CNN model then classifies the detected noisy signal data as originating from either precessing or non-precessing (aligned/anti-aligned) systems. Using simulated data, the trained classifier distinguishes between noise and noisy GW signals with more than 99\% accuracy. The second classifier further differentiates between aligned and highly precessing signals with around 95\% accuracy.
We extended our analysis to a multi-detector framework by performing a coincident test. Additionally, we tested the performance of our trained architecture on data from the first three observation runs (O1, O2, and O3) of LIGO and Virgo to identify detected BBH events as either aligned or precessing.
\end{abstract}
\maketitle
\section{Introduction}
\label{Sec:Intro}
The Advanced LIGO ~\cite{aasi2015advanced} and Virgo ~\cite{acernese2014advanced} interferometric GW detectors have detected $\sim 90$ compact binary sources (comprising of stellar and intermediate-mass black holes and neutron stars) in the first three observational runs ~\cite{gwtc3}. The first phase of the fourth observational (O4a) was conducted by the LIGO detectors during May 24, 2023 to January 16, 2024. Virgo did not join the O4a. Second phase of O4 (O4b) began on April 10, 2024 and is expected to end in June 2025. Virgo has joined the two LIGO detectors in O4b. KAGRA\cite{kagra}, a GW detector in Japan, is also expected to join LIGO and Virgo, towards the end of O4b. 

Most the detected gravitational wave sources reported in 3rd catalog GWTC-3\cite{gwtc3} from the first three observing runs are binary black hole systems. Three events are from neutron star binaries and one from neutron star - black hole binary. The catalog also contains some exceptional events, such as GW190521 \cite{abbott2020gw190521, estelles2022detailed}, GW190412 \cite{rodriguez2020gw190412}, $\text{GW}190915\_{235702}$ ~\cite{hoy2022evidence}, expected to be coming from the precessing binaries. Fourth observing run has reported even more exotic source of gravitational wave $\text{GW}230529\_{181500}$ where companion of a neutron star is an object from the mass gap ($3 - 5 M_{\odot}$)\cite{O4a}.

\begin{figure*}[tbh]
\centering
\includegraphics[width=0.7\linewidth]{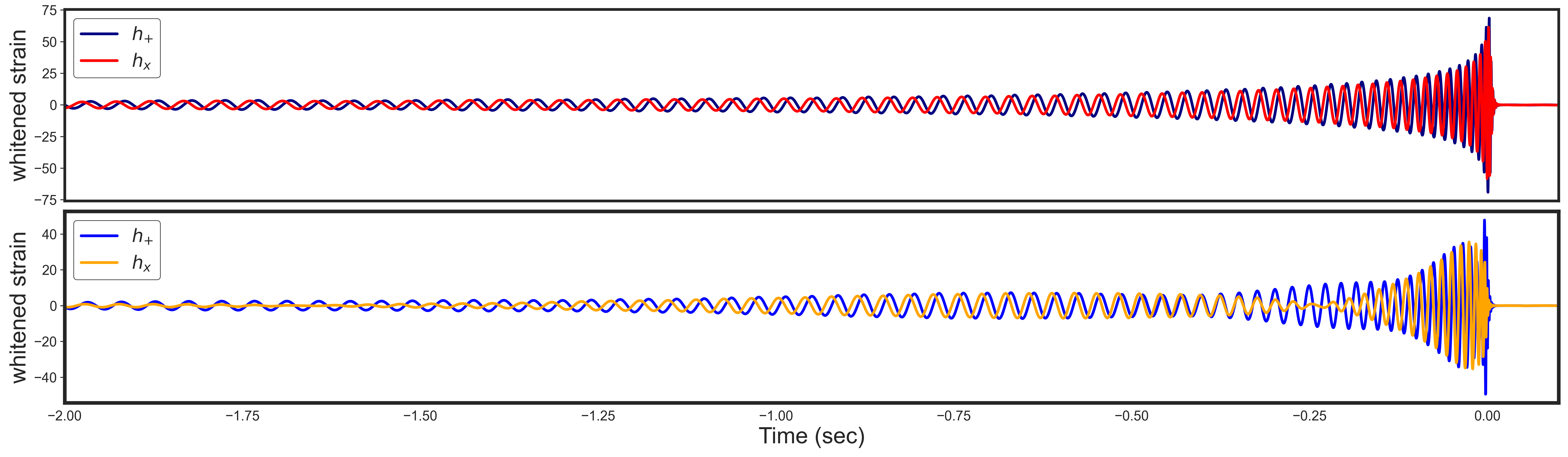}
\caption{The figure ( upper and lower panel ) shows the plus ($h_{+}$ ) and cross ($h_{\times}$)-polarization waveforms generated using $\IMRPhenomD$ and $\IMRPhenomPv$ ~\cite{khan2016frequency, khan2019phenomenological, bustillo2021confusing} waveform models and whitened using a modelled power spectral density $(\PSD)$ ~\cite{moore2014gravitational}. The same mass ($m_{1, 2} = (28.06, 11.29)$) and spin-$z$ components ($s_{1z, 2z} = (-0.11, 0.62)$ ) are chosen to generate aligned and precessing waveforms. 
Binary Black hole spin precession is characterized using a single effective parameter, denoted as $\chi_{\text{p}}$ ~\cite{PrecessingSpin, precessionSNR}. The non-zero $\chi_{\text{p}}$ value indicates the precessing spins. The high value ($ \chi_{\text{p}} \rightarrow 1$) and low value ($ \chi_{\text{p}} \rightarrow 0$ ) would define strong and weak precessing signals. For this specific example, the corresponding $\chi_{\text{p}}$ for the precessing signal is $0.64$.}
\label{fig:waveform}
\end{figure*}

Most black holes are considered to be spinning, often with arbitrary orientation. Neutron stars, on the other hand, are not expected to have significant spin angular momentum. The spins of the black holes in a binary system could be aligned or misaligned with the orbital angular momentum. Systems in which the orbital and spin angular momenta are not aligned are known as precessing systems. A confident detection of a signal from a precessing system would allow us to better understand the formation and evolution of such systems. By measuring the effects of spin precession, we can obtain better measurements of the angular momentum, which can help us disentangle various degenerate parameters (e.g., mass ratio, inclination angle, and angular momentum) ~\cite{Cutler1994, Poisson1995, Emily2013, Salvatore2018, Usman2019constraining, pratten2020measuring}. However, the current search pipelines ~\cite{Cody, usman2016pycbc, MBTA, SPIIR, cWB} for compact binaries are designed to detect only aligned (or anti-aligned) systems and hence do not differentiate between aligned and precessing systems. Thus, the pipelines could miss the precession effects ~\cite{harryPrecession}. The effects of spin precession on their waveforms appear as amplitude and phase modulations (see Figure \ref{fig:waveform}).

Current searches for GW signals from compact binary objects use matched-filter-based detection statistics to identify signals. The matched filter convolves detector data with theoretical waveforms from a discrete set known as a template bank. The bank is constructed so that any signal with parameters within the parameter space retains at least 97\% of its signal-to-noise ratio (SNR) due to a match with a template waveform. In practice, this criterion is valid if the GW signal closely matches the waveform models. By setting a threshold on SNR, the pipeline obtains triggers independently for each detector and uses them for further analysis. The pipelines also employ the coincident method / test by using all possible detector combinations (double or triple) to get coincident triggers. This coincident test is a powerful tool for reducing the number of false triggers originating from instrumental and environmental glitches, and hence the false alarm rate (FAR).

Current pipelines ~\cite{Cutler1994, Poisson1995, Emily2013, Salvatore2018, Usman2019constraining} are limited to performing searches using aligned-spin waveforms. The simple relationship between the orientation of the aligned-spin system and the sky position allows us to maximize the extrinsic parameters analytically for the matched filter (SNR calculation) definition \cite{Findchirp}. However, the same analytical maximization over extrinsic parameters is not possible for a misaligned (precessing) system. Efforts have been made to obtain the matched filter statistics for precessing cases ~\cite{harryPrecession, harry2018searching}. Additionally, searching for precessing signals requires template banks of higher dimensions (e.g., x- and y-components of spins). Thus, the overall volume of the template bank increases, and more templates are required for the signals from precessing searches compared to an aligned search. Consequently, the computational cost of the precessing signal search increases by several orders of magnitude.

Previously proposed methods for developing precessing matched-filter searches have focused on generalizing detection statistics ~\cite{harryPrecession} or utilizing signal decomposition techniques ~\cite{precessionSNR}. In this paper, we present a new search algorithm based on a deep-learning scheme to detect GW signals from compact binaries with precessing spins. GW signal detection can be treated as a classification problem in machine learning (ML), where strain data is classified as either signal or pure noise in a fraction of a second. This makes ML approaches attractive for the low-latency GW searches needed for electromagnetic (EM) follow-up. However, current low-latency matched-filter searches ~\cite{usman2016pycbc, Cody} are limited to non-precessing compact binaries.

Recently, various deep-learning-based methods have been applied for the classification of GW signals from noisy data ~\cite{krastev2020real, lin2020binary, wang2020gravitational, gabbard2018matching, george2018classification, George:2016hay, George:2017pmj, gebhard2019convolutional, li2020some, wei2020gravitational, miller2019effective, ma2022ensemble}. However, to the best of our knowledge, no ML approach has been developed to detect signals from precessing systems. This work represents the first attempt to classify GW signals as either aligned or precessing. While current classical search pipelines can detect signals with small amounts of precession using aligned template waveforms, highly precessing signals can be missed by the template banks employed in these searches. Therefore, it is crucial to explore alternative methods to determine if ML can provide a solution for detecting highly precessing signals.

The paper is organized as follows. In Section \ref{Sec:methods}, we first describe the details of the proposed model. This section presents the architecture used, the data preparation process, and the training and evaluation scheme employed for this analysis. Section \ref{Sec:Evaluation} follows, detailing the performance analysis of the proposed model. Performance evaluation includes testing on specific data chunks and continuous data streams to extract trigger information. In Section \ref{Subsec: Multi-detector}, results are obtained by applying the coincident test for multi-detector scenarios, which helps constrain the event's merger time range. In Section \ref{Sec:RealEvents}, the proposed scheme is applied on real events from the first three observing runs of two LIGO detectors. Finally, Section \ref{Sec:Conclusion} provides a discussion on the outcomes of our studies and concluding remarks.
\section{Methodology}
\label{Sec:methods}
In this study, we focus on detecting GW signals from merging pairs of black holes using Convolutional Neural Networks (CNN). The same methodology applies to GW signals from merging pairs of neutron stars or neutron star-black hole pairs.

The spins of the black holes can be aligned or anti-aligned with the orbital angular momentum, or they may precess around it. We conducted several case studies to evaluate the detection feasibility of GW signals from precessing black hole systems using CNN. The initial study involves binary classification between noisy GW signals and pure noise. GW signals correspond exclusively to the precessing binary black hole systems. A noisy time series, sampled from analytical advanced LIGO power spectral density, is added to simulate noisy GW signals. Subsequently, we extended this study to include signals from precessing and aligned systems.

Another case study involves hierarchical binary classification. Here, two CNN models were employed sequentially: the first model distinguishes pure noise from noisy GW signals, while the second model discerned noisy aligned from noisy precessing GW signals. Alternatively, this classification problem could be framed as a multi-label classification, where pure noise, noisy GW signals from aligned systems, and noisy GW signals from precessing systems are treated as distinct classes. In Section \ref{Sec:Evaluation}, we demonstrate that the hierarchical binary classifiers slightly outperformed the multi-label classifier. Therefore, our reported results are produced using the hierarchical binary classifier.

All case studies utilized the same convolutional neural network (CNN) model described in Section \ref{sec:CNN}. This specific CNN model demonstrated higher accuracy in classifying pure noise versus noisy GW signals than distinguishing between aligned and precessing signals. However, future research could explore the potential benefits of designing different CNN architectures for multi-label or secondary binary classifications between aligned and precessing signals.

\begin{figure}[ht!]
\centering
\includegraphics[scale = 0.25]{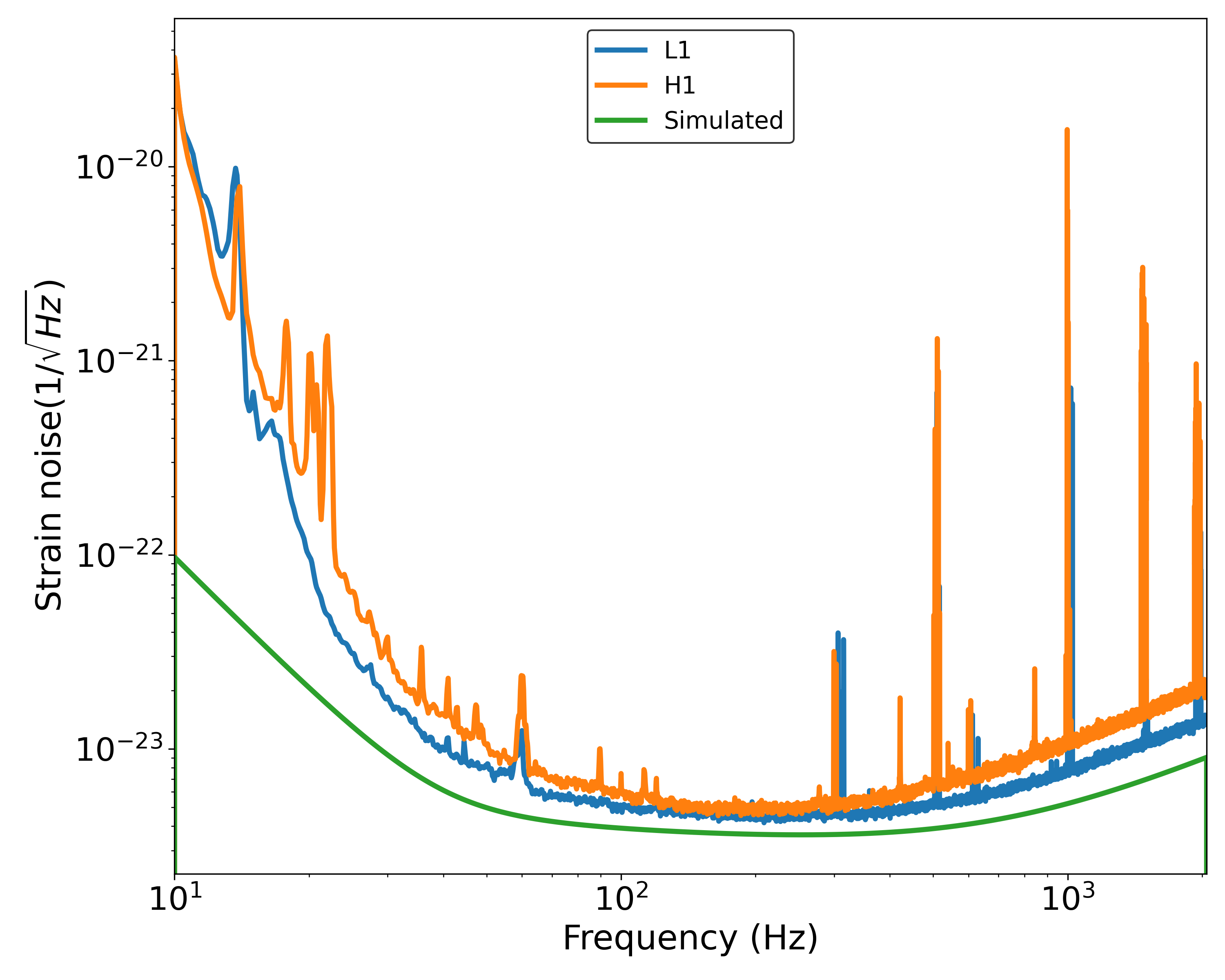}
\caption{This figure depicts the amplitude spectral density(ASD) estimated from O3 data of  two detectors($\text{L}1$ and $\text{H}1$) and simulated ASD using a modelled PSD \cite{moore2014gravitational}.  }
\label{fig:PSD}
\end{figure}

\vspace{.2 in}
\textbf{Strain Data Representation:}
In general, the detector output $d(t)$ is a gravitational wave signal $h(t)$ buried in detector noise $n(t)$: 
\begin{equation}
d(t) \equiv n(t) + h(t) \, ,
\end{equation}
We train our CNN architecture independently with real and simulated data. To model the GW signal $h(t)$, we use the $\text{IMRPhenom}$ (Inspiral-Merger-Ringdown Phenomenological) waveform models ~\cite{khan2016frequency, khan2019phenomenological, bustillo2021confusing} for non-precessing (aligned/anti-aligned) and precessing systems (see Figure \ref{fig:waveform} for a typical structure of the waveforms). The simulated noise has been generated using a modelled $\PSD$ ~\cite{moore2014gravitational}, designed for the advanced LIGO configuration. While the real noise is sampled from the PSD estimated from the first three months of $\O3$ data \cite{aligoO3L1PSD, aligoO3H1PSD} (See Figure \ref{fig:PSD}) of LIGO's Hanford and Livingstone detectors H1 and L1, respectively. The simulated and real noise are added to the signal to generate simulated and real training data, respectively.

We generate waveforms by varying individual masses and the spin parameters $\chi_{\text{eff}}$~\cite{EffectiveSpin} for aligned case and $\chi_{\text{p}}$~\cite{PrecessingSpin, precessionSNR} for precessing case over the ranges outlined in Table \ref{tab:param_boundary}.
$\chi_{\text{eff}} = \frac{m_{1} \chi_{1\parallel} + m_{2} \chi_{2\parallel}}{M}$ is a single mass-weighted effective, aligned spin parameter that combines spins for binary systems \cite{EffectiveSpin, gwtc3}. $\chi_{1 \parallel}, \chi_{2 \parallel}$ are the component spin magnitudes of a binary, $m_{1}$, $m_{2}$ are the component masses of a binary system with $m_{1} > m_{2}$ and $M = m_{1} + m_{2}$ is the total mass of the system.
$\chi_{\text{p}} = \frac{1}{\text{A}_{1} m_{1}^{2}}~\text{max}(\text{A}_{1} m_{1}^{2} \chi_{1\perp}, \text{A}_{2} m_{2}^{2}\chi_{2 \perp})$ describes the precession of a signal. 
$\text{A}_{1} = 2 + \frac{3q}{2}$ and $\text{A}_{2} = 2 + \frac{3}{2q}$. $\chi_{1\perp}$ and $\chi_{2\perp}$ are the in-plane spin magnitudes of component black holes. $\chi_{1\parallel}$ and $\chi_{2\parallel}$ are the spin magnitudes of two black holes parallel to the direction of their orbital angular momenta and $\text{q} = m_{1}/ m_{2}$ represents the component mass ratio.
High $\chi_{\text{p}}$ values imply high precession. However, overall precession depends on other parameters, such as orientation of the binary, i.e, total angular momentum with respect to the line-of-sight, denoted by $\theta_{\text{JN}}$ and the opening angle between total angular momentum and orbital angular momentum, denoted by $\iota$ \cite{precessionSNR}. The orientation $\iota = \theta_{\text{JN}} = 0$ is referred to as `face-on' whereas $\iota = \theta_{\text{JN}} = \pi/2$  is known as `edge on' \cite{Schmidt2014StudyingAM}.
\begin{table}[!ht]
\begin{center}
\begin{tabular}{|c|c|c|c|c|c|c|}
\hline
\textbf{Signal} & \textbf{System} & \bf{$\text{m}_{1} (\text{M}_{\odot})$} &
\textbf{$\text{m}_{2}(\text{M}_{\odot})$} & $q$ &
\textbf{$\chi_{\text{eff}}$} & \textbf{$\chi_{\text{p}}$} \\
\hline
\multirow{2}{*}{$\BBH$} & $\AS$ & [$5$, 95] & [5, 95] & [1, 5]& [0.1, 0.9] & -- \\
&$\PS$ & [5, 95] & [5, 95] & [1, 5] & -- & [0.1, 0.9] \\
\hline
\end{tabular}
\caption{The table describes the intrinsic parameter space that is used to train and test our $\CNN$ architecture. $\text{m}_{1}$, $\text{m}_{2}$ are the component masses of a binary system with $\text{m}_{1} > \text{m}_{2}$. $\text{q}$ represents the component mass ratio. The parameters $\chi_{\text{eff}}$ and $\chi_{\text{p}}$ represents effective spin of aligned spin and precessing spin systems respectively. $\AS$ and $\PS$ represent the aligned and precessing spins systems respectively}
\label{tab:param_boundary}
\end{center}
\end{table}

\textbf{Signal-to-Noise Ratio (SNR):} 
Matched filtering \cite{Findchirp} is an optimal algorithm to search for a known signal in the presence of Gaussian noise. Mathematically it can be expressed as follows:
\begin{equation}
\langle s(t), h(t) \rangle = 4 \mathbb{Re} \,[ \int_{f_{\text{low}}}^{f_{\text{high}}} {\frac{\tilde{s}(f) \, \tilde{h}^{*}(f)}{S_{n}(f) } \, df } \,] , 
\label{Eq:SNR}
\end{equation}
where $\text{S}_{n}(f)$ is defined as the one-sided $\PSD$ and $\mathbb{Re}$[.] denotes the real part. $f_{\text{low}}$ and $f_{\text{high}}$ denote the lower and upper cut-off frequencies in the detector band-width. $\tilde{h}$ represents the GW template (an analytical waveform in frequency domain corresponding to a specific point in the template bank). The square root of Eq.~\ref{Eq:SNR} is known as Signal to Noise Ratio ($\SNR$). The detector output $s(t)$ may or may not contain the true $\GW$ signal.

Using Eq.~\ref{Eq:SNR}, we can define,
\begin{equation}
\langle h(t), h(t) \rangle = 4 \, \mathrm{Re} \, \left[  \int_{f_{\text{low}}}^{f_{\text{high}}}\frac{\lvert\tilde{h}(f)\rvert^{2}}{S_{n}(f)} \, df \, \right] ,
\end{equation}
$\langle h(t), h(t) \rangle$ defines the matched filter of a template with itself and describes the loudness of a signal in the strain data\cite{Findchirp}. In GW data analysis literature, it is often called optimal $\SNR$ ($\rho^{2}_{\text{opt}}$) \cite{Baltus2021}. For a deep learning-based classification framework, $\rho_{\text{opt}}$ is a valuable parameter to generate the (noisy) data for training / validation / testing purpose. We generate the datasets with $\rho_{\text{opt}} \in [10, 20]$ in the steps of unity.
\subsection{CNN architecture}
\label{sec:CNN}
Previous studies \cite{krastev2020real, lin2020binary, wang2020gravitational, gabbard2018matching, george2018classification, George:2016hay, George:2017pmj, gebhard2019convolutional, li2020some, wei2020gravitational, miller2019effective, ma2022ensemble} have utilized one-dimensional (1D) CNN models to train on gravitational wave (GW) strain data as time series. Typically, 1D CNN models are effective for classifying time series data. In contrast, two-dimensional (2D) CNN architectures are better suited for classifying image data due to their ability to slide kernels along both dimensions (width and length).

We conducted our case studies using both 1D and 2D CNN models, with our results primarily based on the 2D CNN architecture. During our experiments, we found that the 2D CNN model had a faster runtime and provided a nominal improvement in accuracy compared to the 1D CNN model for classifying between noise and noisy signals. The 1D CNN model we used is from Gabbard et al. \cite{gabbard2018matching}, and the 2D model is from Krastev et al. \cite{krastev2020real}.

\begin{figure*}[!ht]
\begin{center}
\includegraphics[scale = 0.5]{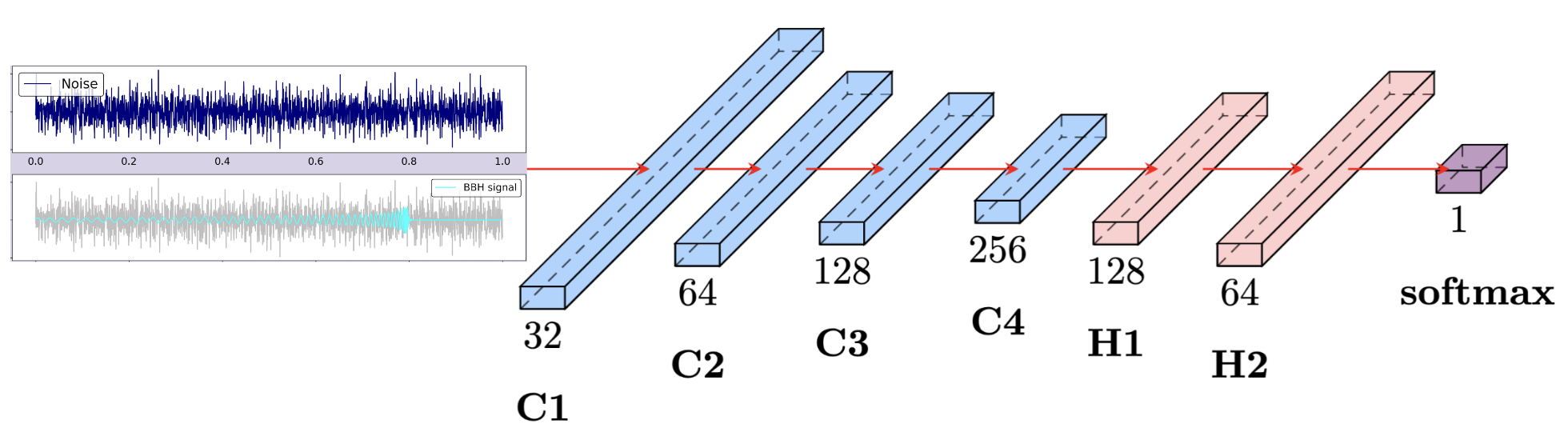}
\caption{
The schematic diagram of CNN architecture for binary classification between pure noise and noisy signal. It contains four convolution layers (C1, C2, C3, and C4), and two hidden layers (HD1, HD2). We used the ``ReLU" activation function for all convolution and hidden layers. However, the activation function: ``softmax" is used for the output layer.
} 
\label{fig:CNN_arch}
\end{center}
\end{figure*}

We maintained the architectural structures (number of convolution and hidden layers) of Gabbard's 1D and Krastev's 2D CNN models but varied the number of epochs, batch size, and learning rate. The number of epochs, which defines the number of complete passes through the training dataset, was varied across different case studies, with a fixed batch size of 32 or 64 and a learning rate of \(10^{-4}\). The specific configuration of our model, including the number of neurons at each layer, activation function, filter size, and hidden layers, is detailed in Table \ref{tab:CNN-arch} and Figure \ref{fig:CNN_arch}.
The ``Binary Cross Entropy" loss function \cite{ho2019binaryCross} is used for the binary classification tasks for the hierarchical classification model. It measures the cross-entropy loss between the predicted and original classes (labels) of the training dataset. For the multi-label classification task, the ``Categorical Cross-Entropy" loss function \cite{zhang2018generalized, rusiecki2019trimmed} is used. 

\begin{table}[hbt]
\centering
\begin{ruledtabular}
\begin{tabular}{l c c c c c c}
Parameter &1 &2 &3 &4 &5 &6 \\
\hline
Type & $\text{C}$ & $\text{C}$ & $\text{C}$ & $\text{C}$ & $\text{H}$ & $\text{H}$\\
Neurons & $32$ & $64$ & $128$ & $256$ & $128$ & $64$ \\
Filter size & ($1$, $16$) & ($1$, $8$) & ($1$, $8$) & ($1$, $8$) & N/A & N/A \\
Maxpool size & ($1$, $4$) & ($1$, $4$) & ($1$, $4$) & ($1$, $4$) & N/A & N/A \\
Drop out & $0$ & $0$ & $0$ & $0$ & $0.5$ & $0.5$ \\
Activation function & $\text{ReLU}$ & $\text{ReLU}$ & $\text{ReLU}$ & $\text{ReLU}$ & $\text{ReLU}$ & $\text{ReLU}$  \\
\end{tabular}
\caption{The $\CNN$ architecture consists of four convolution layers ($\text{C}$) followed by two hidden layers ($\text{H}$). Max pooling is performed at each convolution layer. We also use the dropout layer with a rate of $0.5$ at the hidden layers. The output layer has the number of neurons equal to the number of classes. For the output layer, we used the soft-max ( $\text{S}_{\text{Max}}$) as an activation function, which gives the output in terms of the prediction probabilities. Figure \ref{fig:CNN_arch} depicts the representation of the same CNN architecture for the classification between noisy signal and pure noise.  }
\label{tab:CNN-arch}
\end{ruledtabular}
\end{table}
The number of neurons across the layers is progressively increased to improve the ability of the network to extract more relevant features from the time series data in an initial set of layers, followed by a gradual reduction in the number of neurons in subsequent layers to enable classification. The chosen configuration provides reasonable high accuracy for our training and testing data set. However, new neural network models can be created to obtain similar or better accuracy. 
\subsection{Training Strategies}
The CNN models have been trained with $1.5 \times 10^{5}$ chunks of whitened strain time series data of duration one second and sampling frequency of $4096$ Hz. We have used $\IMRPhenomD$ and $\IMRPhenomPv$ waveform models to generate synthetic signals from the aligned and precessing systems. 
We have adopted distinct training strategies for hierarchical binary and multi-label classification methods. For hierarchical binary classification, which involves two CNN binary classifiers, we divided the training into two parts: 1. a data set consisting of 50\% pure noise and 50\% noisy signals (with an equal number of aligned and precessing signals) for training at the first stage and 2. a data set of 50\% noisy aligned and 50\% noisy precessing signals for training at the second stage. Both binary classifiers are trained using independent datasets, with the datasets generated from the same parameter range highlighted in Table \ref{tab:param_boundary}. The architecture design for these classifiers can be either identical or different. We have used identical architecture for simplicity, though future work could explore the possibility of using two different configurations. Using identical architecture would not affect the classification tasks because the optimal neural weights obtained via back propagation for first-stage classification (pure noise vs. noisy signal) and second-stage classification (aligned signal vs. precessing signal) are entirely different.
Consequently, the number of epochs (i.e. time of training) required to achieve optimal weights would also vary due to the distinct nature of classification tasks. The multi-label classification consists of three classes. The training data set is divided into pure noise, noisy aligned, and precessing signals. 

\begin{figure*}[!ht]
\begin{center}
\includegraphics[scale = 0.4]{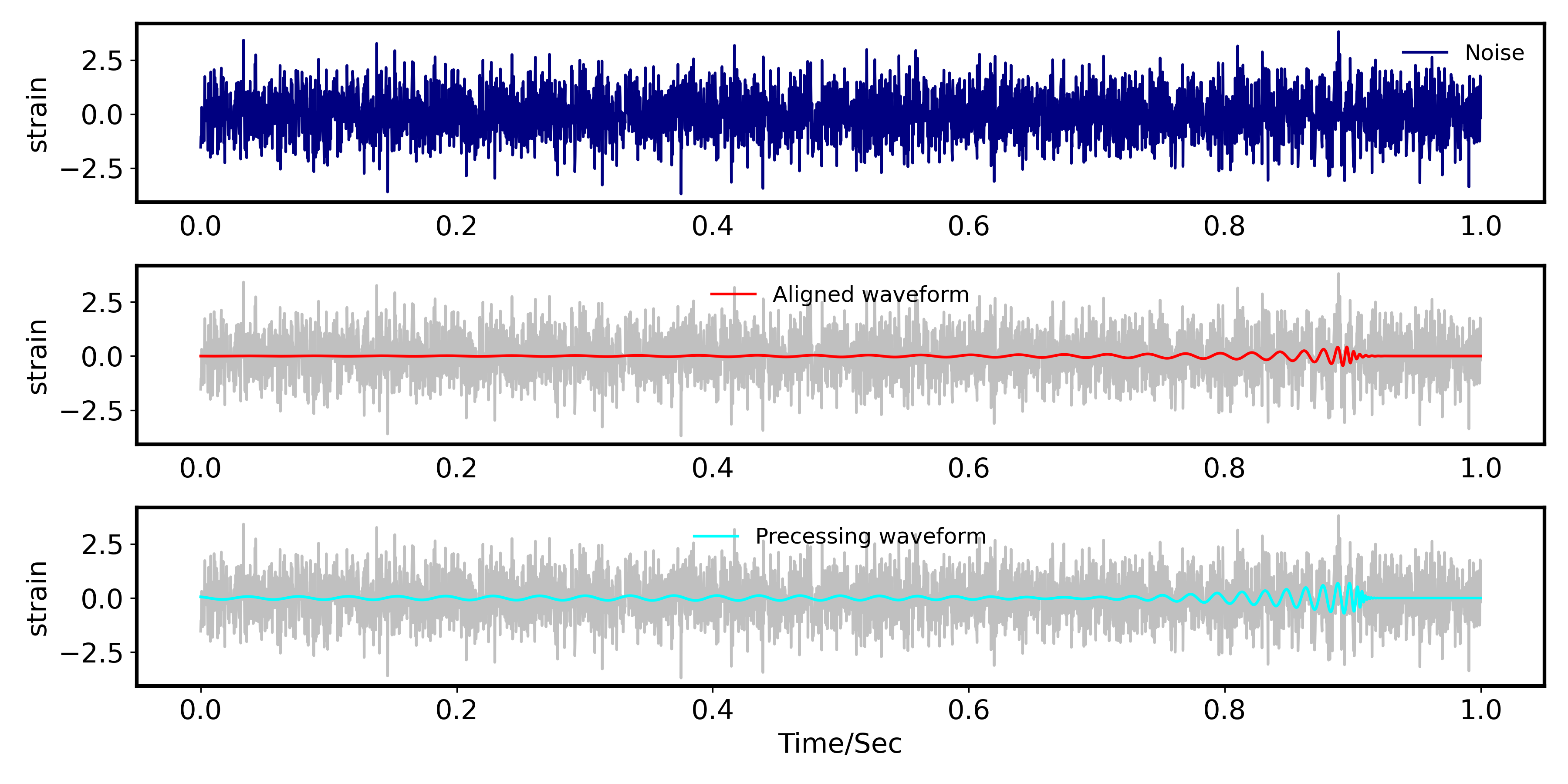}
\caption{The upper panel of the figure shows an example of noisy GW signals. The simulated noise is obtained from the Gaussian distribution. The data is whitened by a modelled $\PSD$ \cite{moore2014gravitational}. The lower panels show whitened strains with $\GW$ signal from aligned and precessing BBH systems. The injected signal from aligned spin $\BBH$ system has component masses $m_{1, 2} = (20.3, 8.8)$ and $\chi_{\text{eff}} = 0.88$. whereas signal from precessing spin system has component masses $m_{1,2} = (30.5, 3.5)$, $\chi_{\text{eff}} = 0.85$ and $\chi_{\text{p}} = 0.7$.} 
\label{fig:Model-I}
\end{center}
\end{figure*}

The component masses for each binary range from $5 M_{\odot}$ to $95 M_{\odot}$, drawn from a uniform distribution. Spins of each black hole in the binary system are also drawn from a uniform distribution such that the parameters $\chi_{\text{eff}}$ and $\chi_{\text{p}}$ both range between $0.1$ and $0.9$ for aligned and precessing spins system, respectively. The extrinsic parameters of each signal, such as right ascension, declination, polarization angle, phase, and inclination angle, are the same as given in \cite{krastev2021detection}. 
The signals are injected in a one-second noise time series such that the signal's merger (peak) time lies between $0.9$ to $0.91$ sec. The injected signals have $\SNR$ between 10 and 20. There are two ways to get the $\SNR$ to a desired range. (a) As $\SNR$ is inversely proportional to the distance of the sources, one can fix a specific distance range to fix the $\SNR$ range. (b) Another option would be to scale the injected normalized signals with the desired $\SNR$. The optimal $\SNR$ is used to normalize the injected signals.
For training and testing, the whitened time series are required. The whitening process involves coloring the time series using the estimated noise $\PSD$. A typical training dataset containing whitened strain data with pure noise and with aligned and precessing waveforms is shown in Figure \ref{fig:Model-I}. To generate the simulated events for testing purposes, we have used Gaussian noise colored by advanced $\LIGO$'s zero-detuned high-power $\PSD$. Whereas, for analyzing the real events from the first three observation runs (O1, O2, and O3), we have colored the Gaussian noise by the O3 $\PSD$s ~\cite{aligoO3H1PSD, aligoO3L1PSD}.
\subsection{Performance Evaluation of the Classifiers}
\label{Sec:Evaluation}
\subsection*{Case Study I: Noise Vs Signal}
As the first case study, we have a CNN model trained to classify between noisy signals and pure noise. To test the performance of this CNN classifier, we prepare $10^{4}$ testing samples comprising of $50\%$ noise and $50\%$ noisy signal samples (divided equally into aligned and processing signals). Each testing sample is a one-second time series whitened by the noise $\PSD$, simulated as well as real. 
The classifier classifies signals from noise with more than $99\%$ accuracy for both kinds of noise.

As a pre-requisite of this task, we have first explored the binary classification between pure noise and noisy precessing signals, and the CNN model successfully predicts precessing signals with more than $99\%$ accuracy. 
\subsection*{Case Study II: Noise Vs Aligned Vs Precessing - Hierarchical Binary Classification Treatment}
\begin{figure*}[tbh]
\centering
\includegraphics[scale = 0.35]{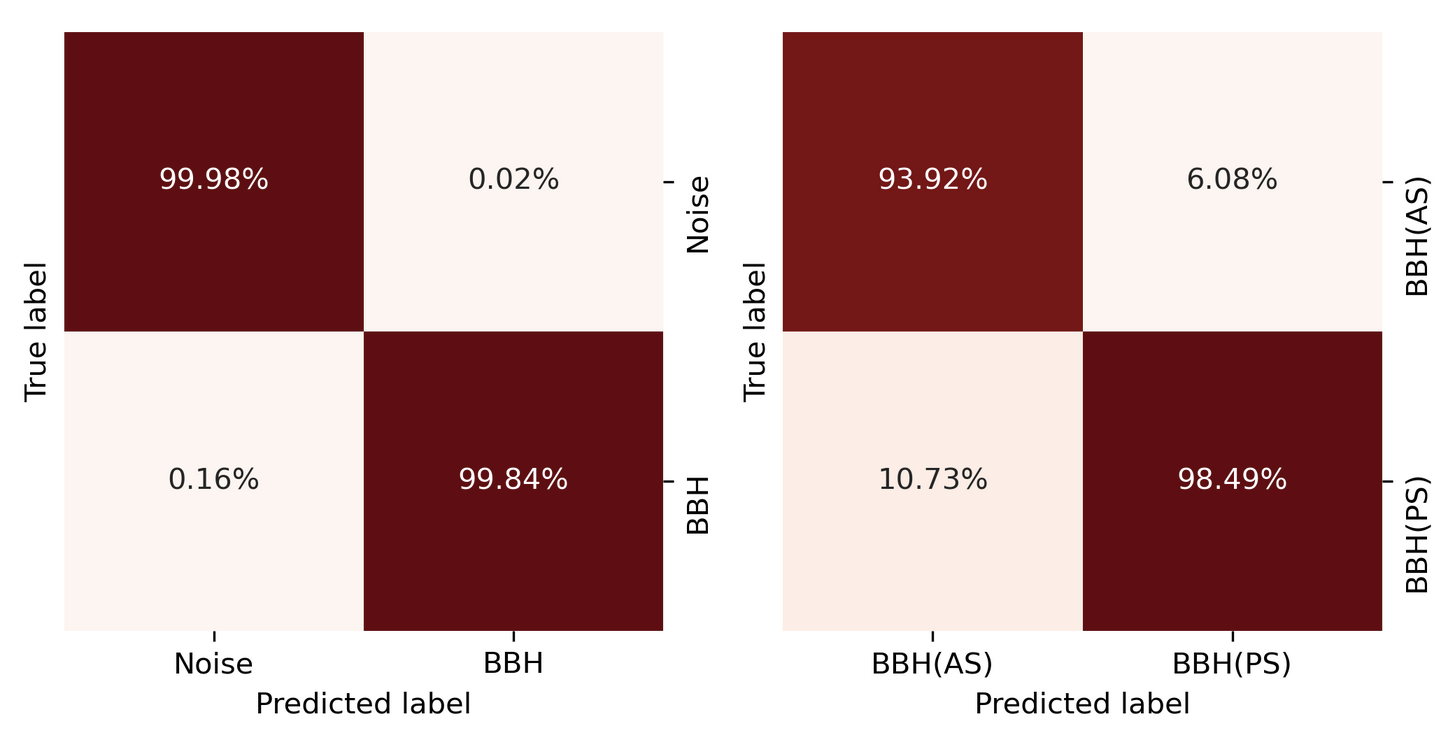}
\caption{The left and right panels show the confusion matrices for the individual CNN models (one for Noise and BBH Signals, another for BBH aligned and precessing spins) for the synthetic data. The training process is independent, but the testing is connected. The classified signals from the first  CNN model are used to test the performance of the second CNN model. }
\label{fig:BBH_GaussNoise_CM}
\end{figure*}

\begin{figure}[!ht]
\begin{center}
\includegraphics[scale = 0.3]{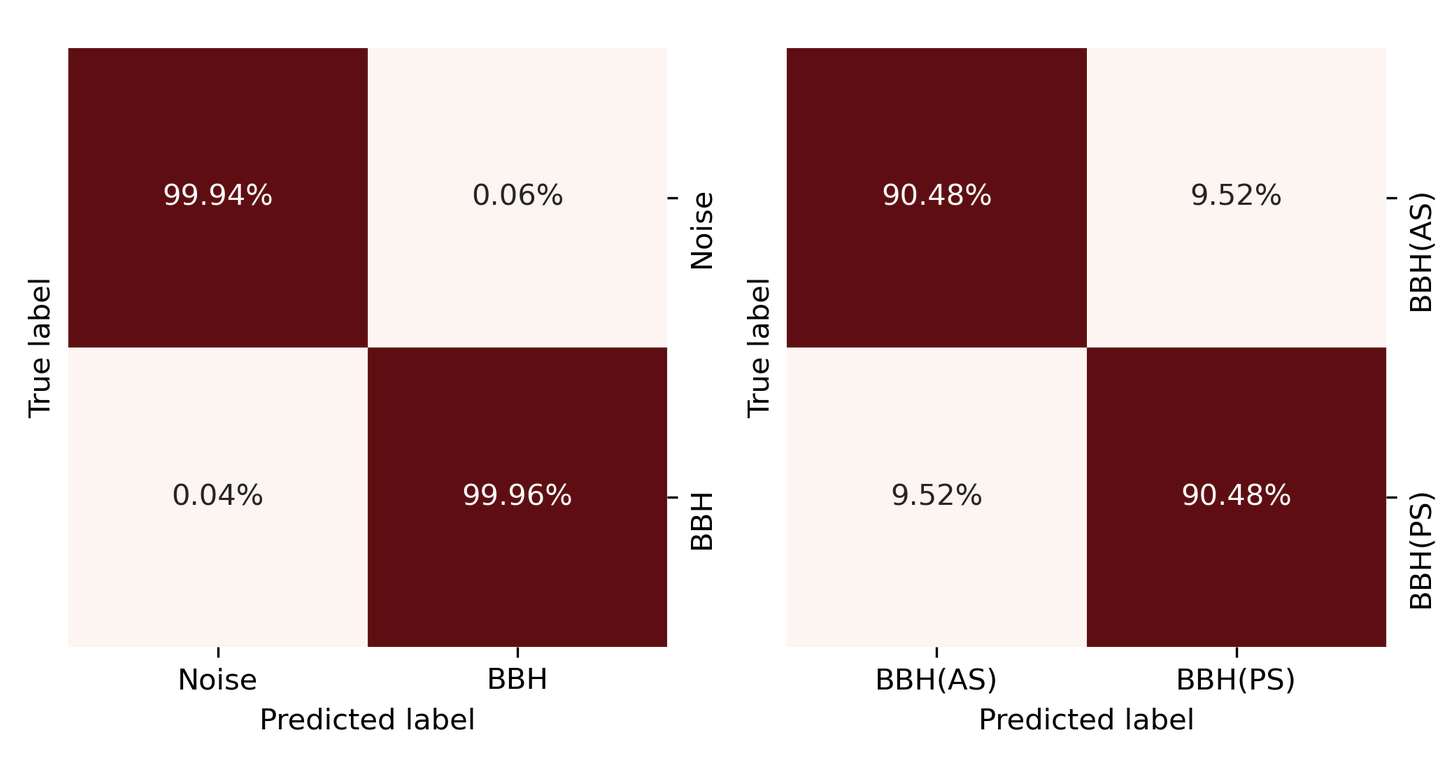}
\caption{
The confusion matrices for the hierarchical binary classification scheme use H$1$ detector noise to generate pure noise and noisy signals (aligned and precessing).
} 
\label{fig:BBH_H1Noise_CM}
\end{center}
\end{figure}

\begin{figure}[!ht]
\begin{center}
\includegraphics[scale = 0.15]{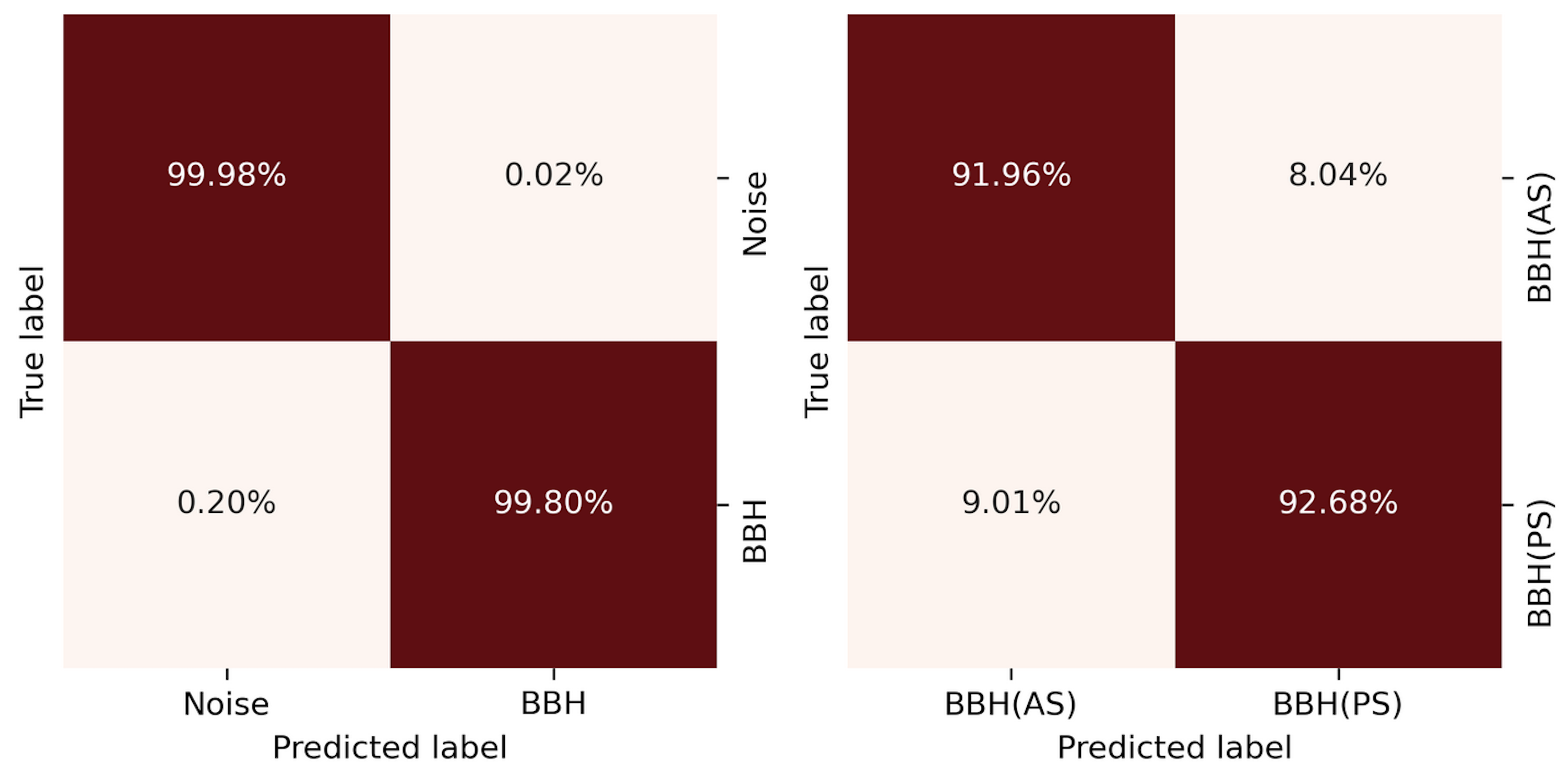}
\caption{
The confusion matrices for the hierarchical binary classification scheme using $\text{L}1$ detector noise.
} 
\label{fig:BBH_L1Noise_CM}
\end{center}
\end{figure}

As mentioned earlier, the training of the classifiers in this case occurs in two independent stages, but their testing is interconnected. The first classifier distinguishes noisy signals from noise with over 99\% accuracy for both types of noise. The correctly classified noisy signals from this stage are then passed to the second classifier to categorize them into aligned and precessing signals. Results indicate that the second classifier achieves an average prediction accuracy of approximately 96.2\% for simulated noise and 90.48\%, 92.32\% for H1, L1 noise respectively. This indicates an overall accuracy, of the two stages combined, to be ~ 97.6\%, 94.74\% and 95.66\% for simulated, H1, and L1 noise, respectively. We performed a 10-fold cross-validation to obtain these classification accuracies.
The performance of the model for Gaussian, real ($\text{H}1$ and $\text{L}1$) noise is shown in Figures \ref{fig:BBH_GaussNoise_CM}, \ref{fig:BBH_H1Noise_CM} and \ref{fig:BBH_L1Noise_CM}, respectively. 
\subsection*{Case Study III: Noise Vs Aligned Vs Precessing - Multi-label Classification Treatment}
\begin{figure}[!ht]
\begin{center}
\includegraphics[scale = 0.3]{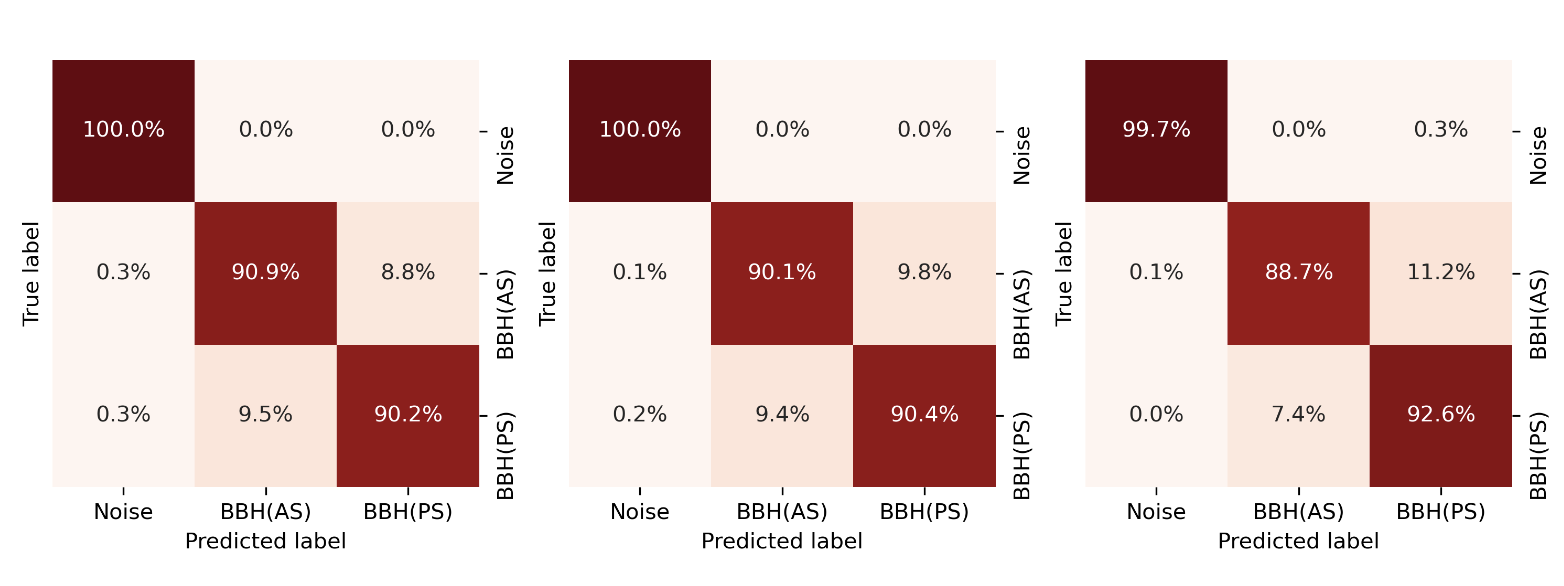}
\caption{Confusion matrices for multi-label classification for synthetic and real data from H$1$ and L$1$. 
} 
\label{fig:all_cases_CM}
\end{center}
\end{figure}

As mentioned in the previous section, the problem of distinguishing between aligned and precessing signals can also be formulated as a multi-class classification problem. In this case, instead of training two independent neural network models, we train one model only with three classes: noise, noisy aligned and noise precessing signals. To evaluate the performance of this multi-class classifier, we feed it with 3000 whitened time series, each of one second long and equally divided into three classes: noise, signals from aligned sources, and signals from precessing sources. The classifier achieves an overall accuracy of 93.7\% for simulated noise, and 93.5\% and 93.6\% for H1 and L1 noise respectively (Figure \ref{fig:all_cases_CM}). This shows that the accuracies obtained from the hierarchical binary classifiers and the multi-class classifier are nearly identical, with the former doing only marginally better. We have chosen the hierarchical binary classifiers method. However, as the performance of both methods is comparable, one can also opt for a multi-class classification method for the same study.
\subsection*{Parameter Space Analysis for Distinguishing Aligned and Precessing Gravitational Waves}
\begin{figure}[!ht]
\begin{center}
\includegraphics[scale = 0.5]{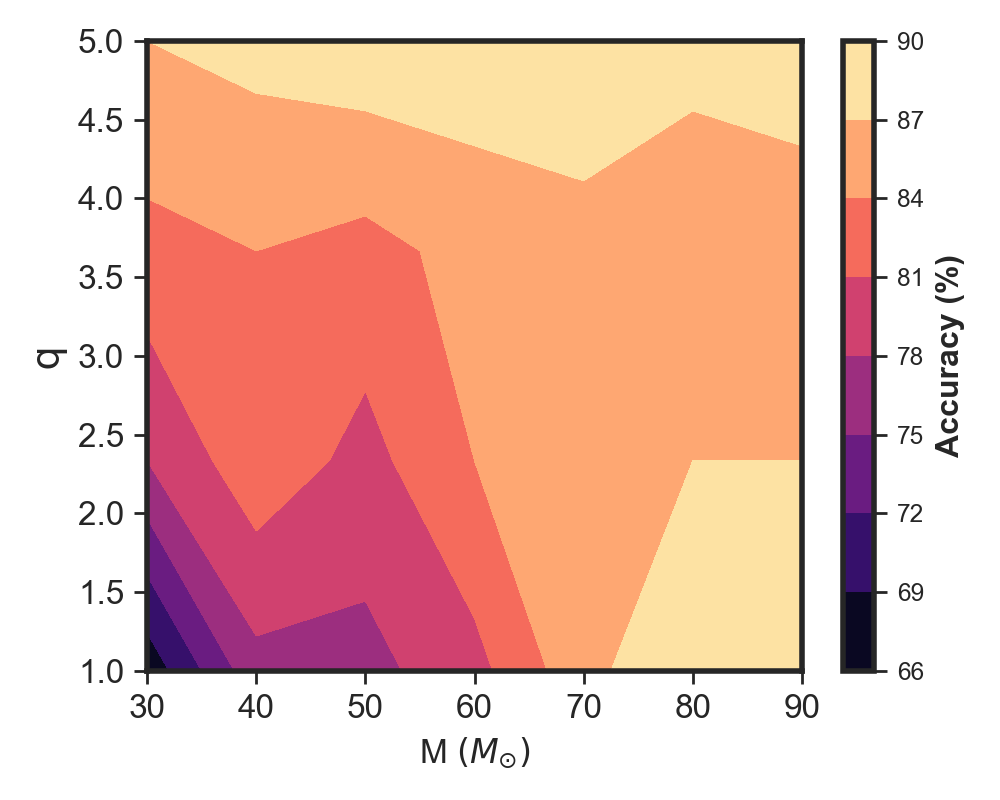}
\caption{
The performance of the trained CNN models to identify the precessing injections (buried in Gaussian noise) is shown across the two-dimensional parameter space by varying total mass ($M$) and mass ratio ($q$). A $5000$ test data set with noisy precessing signals has been created for each bin. The opening angle between orbital angular momentum $L$ and total angular momentum $J$ takes value from $0-60$ degrees while the inclination angle varies from $0-90$ degrees to generate the injections. Correct classification accuracy gradually increases with higher total mass and mass ratio. The difference between aligned and precessing representation is more evident for high total mass and mass ratio regimes. Therefore, our trained model can distinguish such signals with very high accuracy. 
} 
\label{Fig:Acc_M_30_90_q_1_5}
\end{center}
\end{figure}

\begin{figure}[!ht]
\begin{center}
\includegraphics[scale = 0.5]{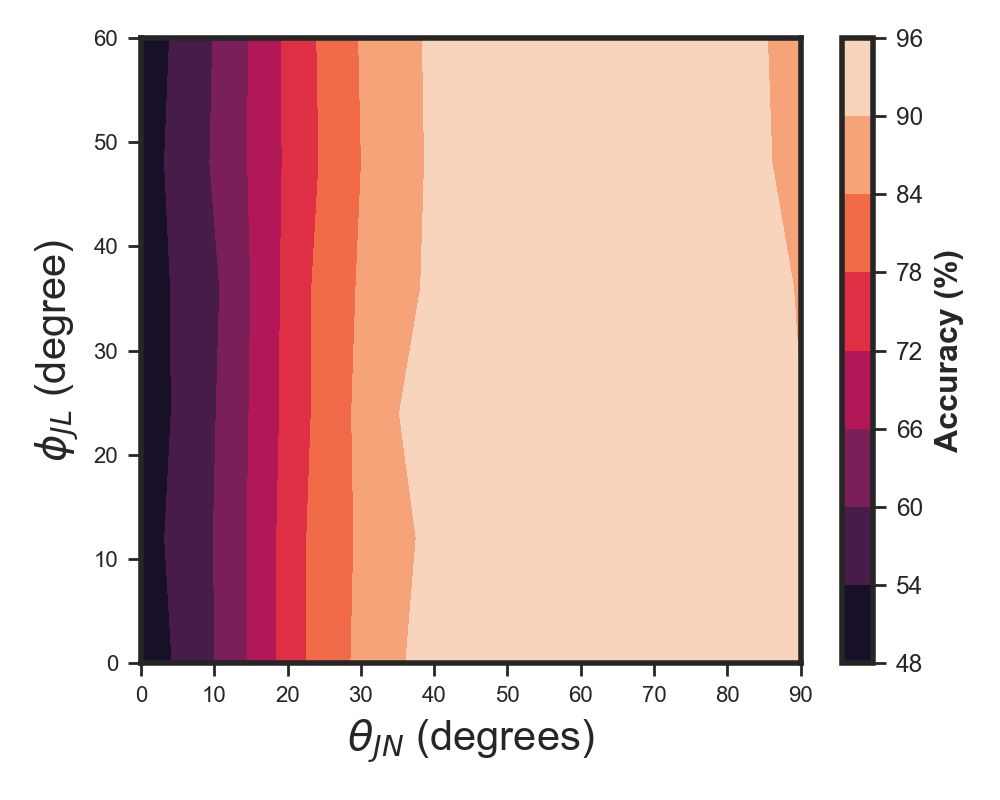}
\caption{Performance of the CNN model to predict precessing signals (embedded in Gaussian noise) in the inclination angle ($\theta_{JN}$) and opening angle ($\phi_{JL}$) space. For each bin corresponding to $\theta_{JN}$ and $\phi_{JL}$, $5000$ precessing signals were simulated. The total mass and mass ratio for signals take the values $30-90 M_{\odot}$ and $1-5$, respectively.} 
\label{Fig:Acc_theta_0_90_phi_0_60}
\end{center}
\end{figure}

From Figures \ref{fig:BBH_GaussNoise_CM}, \ref{fig:BBH_H1Noise_CM}, and \ref{fig:BBH_L1Noise_CM}, it is evident that the classification between pure noise and noisy signals is highly accurate. However, the same level of accuracy is not achieved in the classification between aligned and precessing signals. This suggests that the classifier sometimes struggles to correctly identify whether a signal is from an aligned or precessing system. In cases of low precession, the morphology of aligned and precessing signals may be very similar. Since the CNN architecture learns classification based on the features (morphology) present in the dataset, the model may fail to distinguish between aligned and precessing signals when their features are similar. Therefore, we conducted further investigation to understand the regions of the parameter space where the accuracy is high (with highly precessing signals) and low (with low precessing signals). 
We know that the precession in binary systems depends on several parameters, such as $M$, $q$, $\theta_{\text{JN}}$, and $\phi_{\text{JL}}$. So, we generated several test datasets by varying these parameters. Specifically, we chose total mass and mass ratio ranges between $30-90 M_{\odot}$ and $1-5$, respectively. For each grid point in this range, we generated a test dataset containing 5000 signals, with $\theta_{\text{JN}} \in [0, \frac{\pi}{2}]$ and $\phi_{\text{JL}} \in [0, \frac{\pi}{3}]$. Figure \ref{Fig:Acc_M_30_90_q_1_5} shows the accuracy of our classifiers within this specific parameter space. 
The classification accuracy is high ($\geq 80 \%$) when either the mass ratio or the total mass of binaries is high (e.g., $\text{M} \geq 50 M_{\odot}, q \geq 2$), that shows our model can detect highly precessing systems in such regimes better. Conversely, the accuracy is low for low total mass ($30-50 M_{\odot}$) and low mass ratio ($1-2$), indicating that in this regime, detectability of low precessing systems is more difficult because the model can be confused between aligned and low-precessing signals representation.
In this specific test example (Figure \ref{Fig:Acc_M_30_90_q_1_5}), the highest accuracy achieved is 90\%. This is because $\theta_{\text{JN}}$ and $\phi_{\text{JL}}$ were chosen randomly from a uniform distribution, leading to a mix of signals with moderate or low precession, even in high total mass and mass ratio regimes. Insufficient number of highly precessing signals during the training process can result in such accuracy. To address this, we generated new datasets by varying $\theta_{\text{JN}}$ and $\phi_{\text{JL}}$. Figure \ref{Fig:Acc_theta_0_90_phi_0_60} shows the accuracy for these example datasets. The accuracy of classification increased to 96\% for $\theta_{\text{JN}} \in [40, 90]$ degrees. This test indicates that our classifiers are more sensitive to classifying signals with high precession. The accuracy of our classifiers for low-precession signals is less compared to high-precession signals. This complements the current search schemes searching for precessing systems using aligned template banks. The current search frameworks in LIGO, based on aligned template banks, may not miss signals with low precession but can miss high-precession signals. Thus, a CNN-based classification architecture could be a better alternative for detecting high-precession signals.
We further explored different parameter space configurations to check the consistency of accuracy with high-precession signals. Figures \ref{Fig:Acc_M_q} and \ref{Fig:Acc_theta_phi} show similar accuracies, supporting our findings.

The follow-up section discusses utilizing the trained Neural Network model for real-time detection of GW signals from a multi-detector framework.
\section{Extension to a Multi-detector Case}
\label{Subsec: Multi-detector}
\begin{figure}[!ht]
\begin{center}
\includegraphics[scale = 0.2]{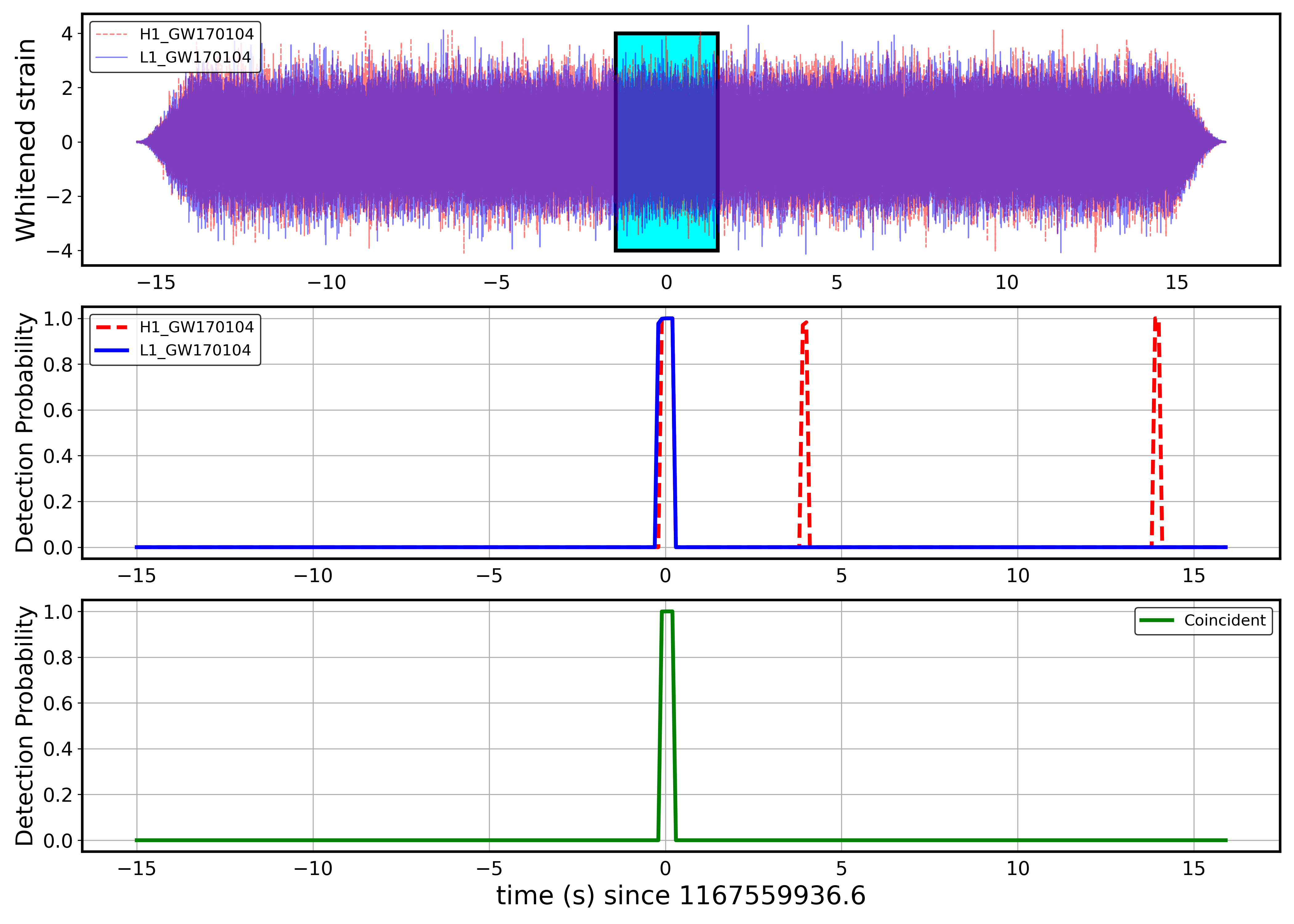}
\caption{The detection of the GW170104 event using the coincident test via the CNN model is shown. The first panel shows the whitened strain data of L$1$ and H$1$ detectors, which are scanned in the $1$ second window (in cyan color) for both the detectors to find the probable GW event. The second panel shows the corresponding detection probability values (outcome of the last layer) for L$1$ (in blue) and H$1$ (in red) detectors. An overlap of detection probability values for both detectors determines the coincident event at the overlapped time region, as shown in the third panel.} 
\label{fig:Coincident-test}
\end{center}
\end{figure}

Incorporation of a coincidence test across multiple detectors would reduce the chance of detecting false signals. Our multiple detector coincident test study is limited to only two detectors (H$1$ and L$1$). Figure \ref{fig:Coincident-test} illustrates the coincidence test scheme applied in this study. 
CNN classifiers only identify the data chunk containing a possible GW signal. Identifying the signal's location in the data chunk is crucial. The coincident test across the detectors is the only alternative to confine the time window in which the signal is presented. We performed the coincident test on our simulated data. We observe many triggers with high detection probability ( based on the softmax values) arising across both the detectors at different locations in the $32$ sec long time series. We notice that the output of H$1$ (in red) and L$1$ (in blue) detectors show a high probability of a trigger at the exact location in time (around $16$ sec) and hence are defined as coincident triggers. It is crucial to mention that the definition of GW triggers obtained from running the classic search pipelines (e.g., PyCBC, GstLAL) differs slightly from the definition of triggers (obtained from CNN classifiers) used in this analysis. GW triggers obtained from classical data analysis pipelines contain information about the intrinsic parameters based on the matched template waveforms. The triggers obtained from CNN classifiers are defined only by the detection probability, and coincident triggers further contain information about the coalescence time. 
In our example, the output of the H$1$ detector shows the triggers having high probability values around $20$ sec and $30$ sec, which are not visible in the output of the L$1$ detector and hence are non-coincident. These non-coincidence triggers are labeled as false alarms. 
\subsection{Adaptation for Real-time Detection of Continuous Time Data}
\begin{figure}[!ht]
\centering
\includegraphics[width = 0.7\linewidth]{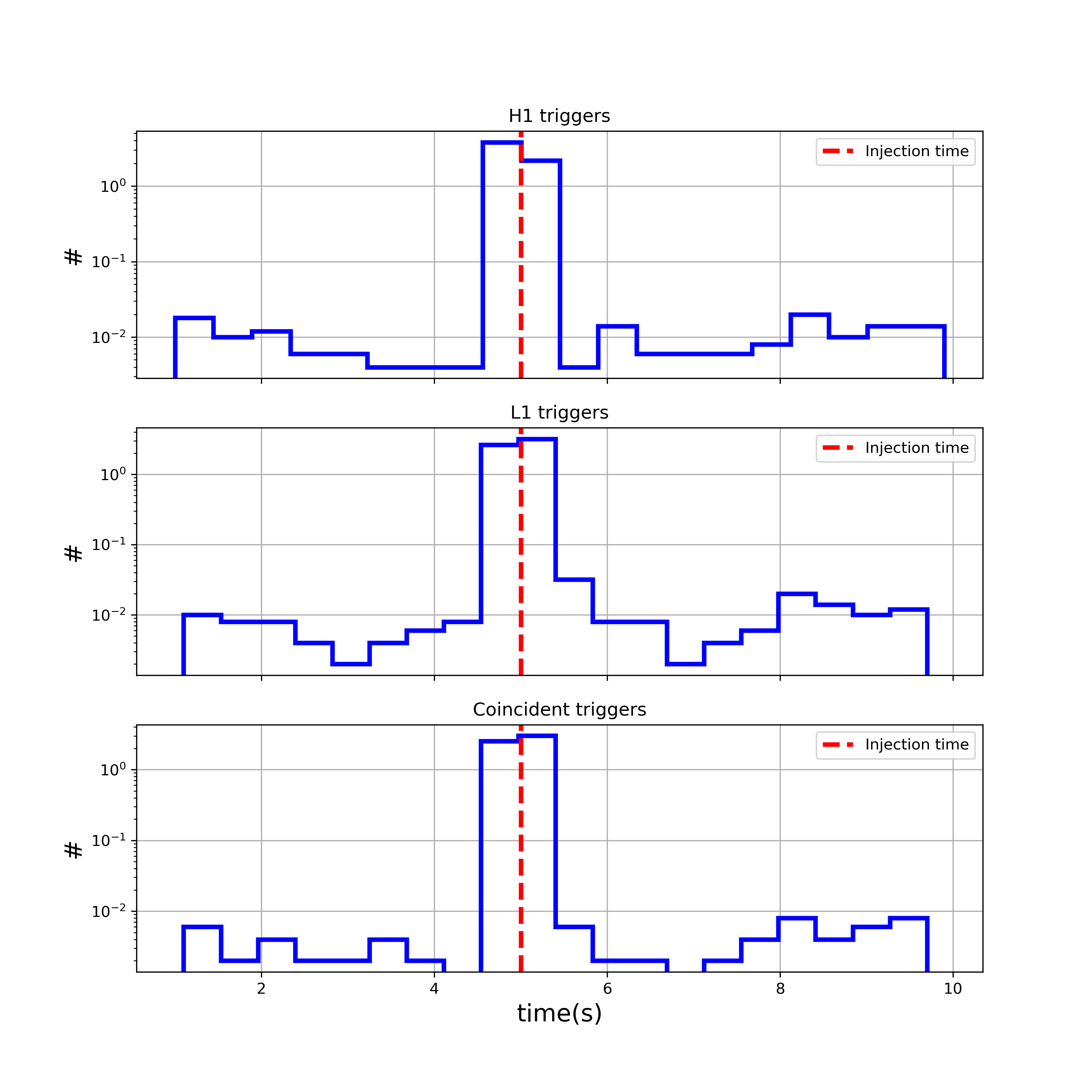}
\caption{The upper, middle, and lower panels show the average distribution of triggers obtained from the two detectors ($\text{H}1$ and $\text{L}1$) noise and that of their coincident output, respectively. The average distribution is obtained by dividing the numbers in each bin by the number of $\BBH$ injections. The red vertical line indicates the merger time of injections. }
\label{fig:coincHist}
\end{figure}

The real detector output is the continuous stretch of time series data. The classical detection pipelines divide the data into small chunks to perform the overlap calculation (match-filtering operations ) between data chunks and a set of analytical waveforms. Taking inspiration from this process, we also tested the performance of our trained model on a continuous stretch of data by dividing it into $10$-sec long time-series chunks. We make $10^{3}$ such chunks (testing time-series samples) of $50\%$  signals and $50\%$ noise for both detectors. In each noisy signal sample, injections ($\text{IMRPhenom}$ waveforms) are placed with the peak position in the middle of the time series, i.e., at $5$ seconds. However, this choice can be varied, and one can randomly place the injections at different time windows. 
As our CNN models are trained with one second-long sample time series ( pure noise and noisy signals), we use a moving window method to analyze the longer duration of data chunks. We place the one-sec window at the beginning of the data chunk ($10$ sec) and shift it by $0.1$ seconds until we reach the end of the data segment. The corresponding probability values to predict a noisy signal or pure noise are recorded with every $0.1$ sec slide. If there is no signal in that segment, we observe a high (higher than a certain threshold) probability value corresponding to the noise class and, consequently, a low (lower than a certain threshold) probability value for the signal class. If a signal is present in the data, as soon as the window starts to overlap with the segment where a signal may be present, we observe the high (low) probability values corresponding to the signal (noise). As we slide the window further, the probability values for the signal start to decrease (increases for noise) as soon as the window recedes away from the signal. For every high probability value at a time step corresponding to the signal, we also record the probability values corresponding to aligned and precessing spin signals obtained at the second stage of the classifier. The average distribution, corresponding to all $10^{4}$ injections, of the triggers in H$1$ and L$1$ data is shown in the first two panels of Figure \ref{fig:coincHist}.

\subsection{Multi-detector Coincidence Test}
\begin{figure}[!ht]
\centering
\includegraphics[width = 0.9\linewidth]{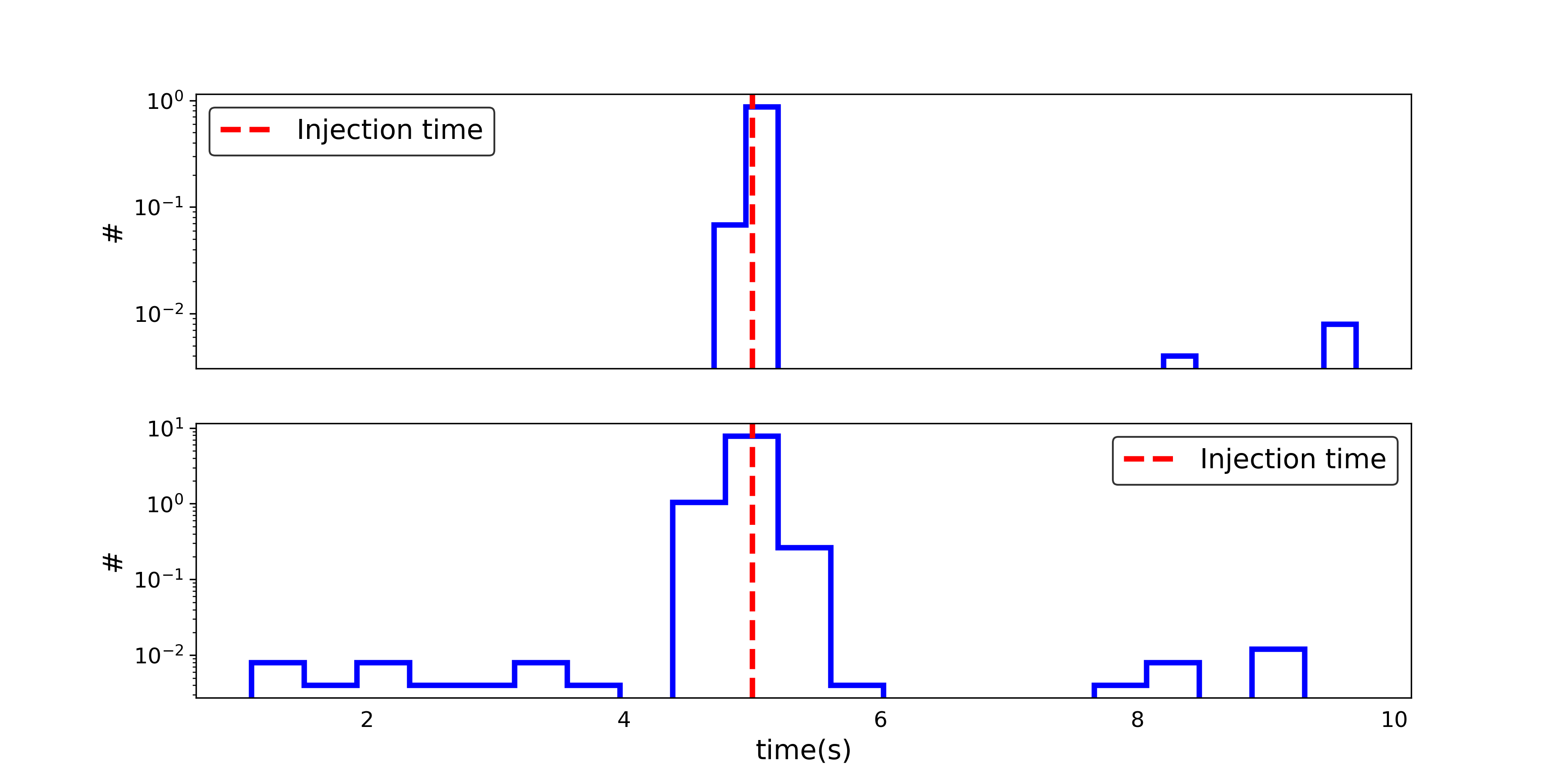}
\caption{The upper panel shows the average distribution of coincident triggers corresponding to the aligned spin injections. The lower penal shows the same for precessing injections. The average distribution is obtained by dividing the numbers in each bin by the number of aligned/precessing injections. The red vertical line indicates the merger time of injections. The threshold of the detection is chosen to be $0.90$.}
\label{fig:ASPSHist}
\end{figure}

After recording the prediction probability values corresponding to the strain data for each detector, we perform the coincidence test on the triggers generated by the two detectors. We define a coincident trigger if we obtain a high probability value for both detectors simultaneously. The third panel of Figure ~\ref{fig:coincHist} shows the average distribution of the coincident triggers across the output of the two detectors. Comparing the first two panels of Figure ~\ref{fig:coincHist} with the third panel, we observe that the false triggers (away from the injection time) reduce significantly in the coincident test. Once we get the coincident triggers from the first stage of the hierarchical binary classification scheme, we perform the same coincident test to find the coincident aligned and precessing triggers. Figure ~\ref{fig:ASPSHist} shows the average distribution of the coincident triggers corresponding to aligned and precessing signals in the two panels. From Figure ~\ref{fig:ASPSHist}, we also observed that, on average, precessing signal generates more triggers around them than an aligned signal.
Our coincident scheme recovers $498$ $\BBH$ signals from $500$ injected at the first stage. Moreover, out of $498$, $214$ signals are recovered as aligned and $206$ as precessing signals at the second stage.
\subsection{Confinement of Merger Time}
\begin{figure}[hbt]
\centering
\begin{tabular}{cc}
\includegraphics[scale = 0.15]{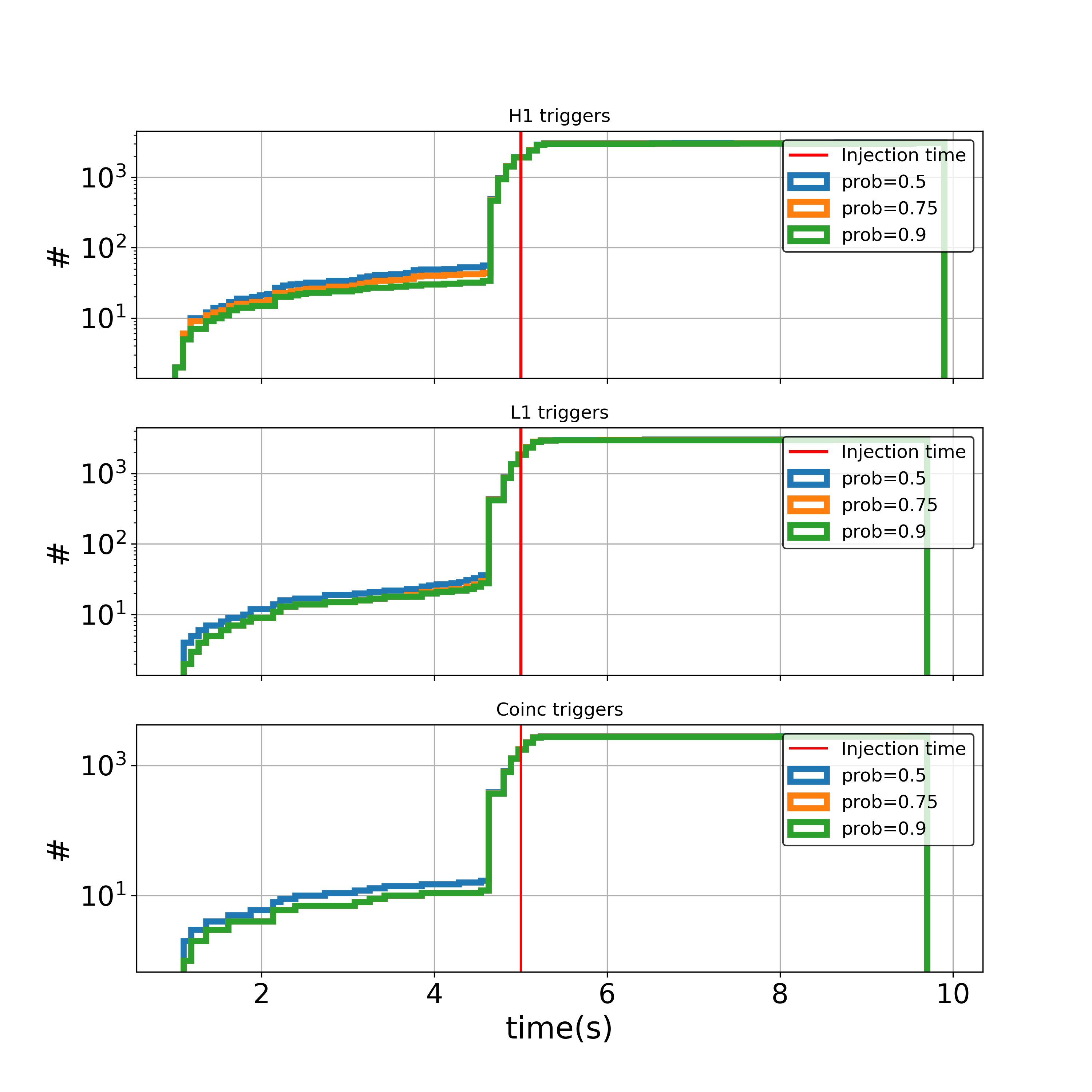}&  
\includegraphics[scale = 0.22]{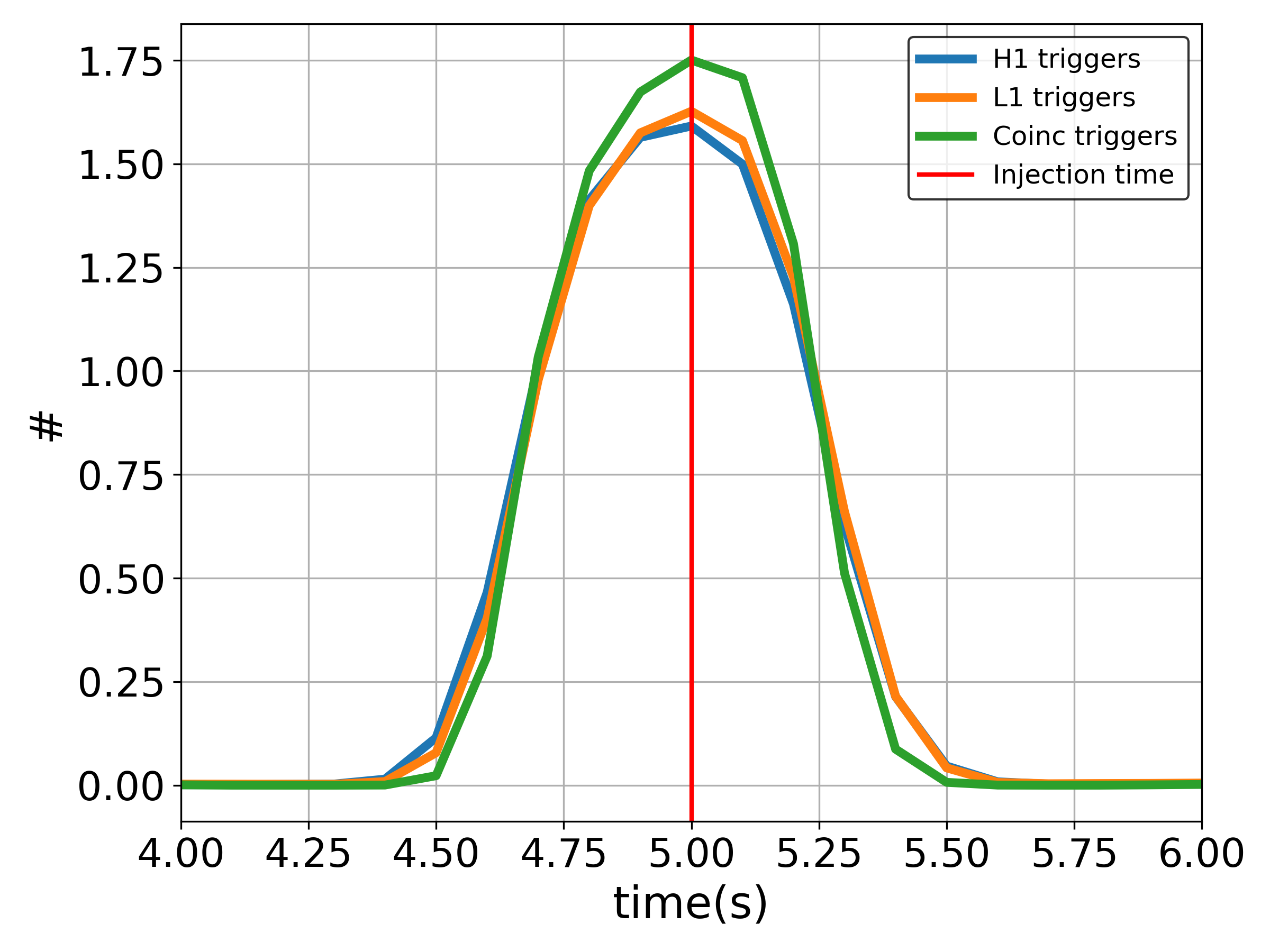}\\
(a) & (b)
\end{tabular}
\caption{(a) The cumulative histogram of H$1$ and L$1$ and the coincident triggers are obtained for different detection probability thresholds($0.50$, $0.75$, and $0.90$). (b) The density plots for the triggers obtained from H$1$ and L$1$ detector noises and their coincidence outcome.}
\label{fig:L1H1-Cum-Coinc}
\end{figure}

In Figure \ref{fig:coincHist}, we obtained the average distribution of the triggers for the individual detectors ($\text{H}1$ and $\text{L}1$) and the coincident triggers for a threshold probability value of $0.9$. We vary the threshold probability of the prediction of a class to observe the variations in the rate of trigger generation. We plot the cumulative histogram of the individual detector triggers and the coincident triggers at different threshold softmax ($0.5$, $0.75$, and $0.9$) (Figure ~\ref{fig:L1H1-Cum-Coinc} (a)). The number of triggers away from the injected time increases with a decrease in the probability threshold, a characteristic of the background triggers. However, the number of triggers around the injection time does not vary too much with a change in the probability threshold, suggesting that the triggers originated from a true event. Hence, these can be considered foreground ground triggers. We plot the density function of individual detectors' triggers and the coincident triggers (Figure \ref{fig:L1H1-Cum-Coinc} (b)). To measure the spread of these distribution of H$1$, L$1$ and coincident triggers, we measure the standard deviations ($\sigma_{\text{H}1}$, $\sigma_{\text{L}1}$ and $\sigma_{\text{Coinc}}$ respectively) associated with them. We find that the spread of the distribution is minimum for the coincident triggers: $\sigma_{\text{Coinc}} = 0.37$ while $\text{H}1$ has the maximum spread($\sigma_{\text{H}1} = 0.54$) and $\text{L}1$ has the intermediate value($\sigma_{\text{L}1} = 0.48$). The mean of the three distributions are $4.98$, $5.00$, and $4.98$. Hence, we found that the coincidence test not only helps to filter out the non-coincident triggers (glitches or false alarms) but also helps confine the merger time of the signals.
\section{Performance on O1, O2 and O3 data}
\label{Sec:RealEvents}
We test the performance of our trained models against the real GW events obtained in the first three observational runs O1, O2, and O3 \footnote{ Data has been taken from GWOSC \url{https://www.gw-openscience.org/data/}} data included in GWTC-1, GWTC-2, and GWTC-3 catalogs. To remind ourselves, we trained our Neural Network model using the sensitivity curves of H1 and L1 based on the first three months of O3 data. Since our training is limited to the optimal SNR range between $10$ to $20$, we only choose those BBH events for which the network $\SNR$ lies in this specific range, specifically network SNR $\geq 13$. Our first binary classifier successfully detects all $\GW$ events as possible signal presence in the data chunk. Figures \ref{fig:ProbGWTC1}, \ref{fig:ProbGWTC2} and \ref{fig:ProbGWTC3} show the corresponding $\GW$ detection probability plots and their coincidence in H1 and L1 detectors. Further, the second CNN model predicts the source of these signals with very high accuracy. The list of correctly classified aligned and precessing events are listed in the tables \ref{tab:GWTC1}, \ref{tab:GWTC2}, \ref{tab:GWTC3}. There is a disagreement between two CNN models dedicated to training on H1 and L1 noise data in classifying them as aligned/precessing signals for some of the events. In those cases, we would depend upon the result of the $\text{L}1$ detector as it is more sensitive than $\text{H}1$.

Readers should note that parameter estimation studies on catalog events do not conclusively determine if the detected events are aligned or precessing, as the confidence intervals are very large — sometimes as large as the width of the priors. However, a few events, such as GW190521\cite{abbott2020gw190521, estelles2022detailed}, GW190412\cite{rodriguez2020gw190412}, and GW190915\_235702\cite{hoy2022evidence}, have been studied in detail as potential precessing signals. Below, we discuss our results on events from GW catalogs with individual SNR $\ge 10$, including those events predicted to be precessing.

\textbf{Events from GWTC-1}
Based on the prediction of our trained classifier, GW150914 is identified as an event from a BBH precessing system, while GW170104 and GW170608 are predicted as events from BBH with aligned spins. For GW170814, the two detectors show a disagreement in their results, leaving us without conclusive evidence to support the nature of this event. None of the GW events in GWTC-1 were reported to exhibit clear precession. However, for GW150914, the reported $\chi_{p}$ based on precessing waveform models (precessing $\text{EOBNR}$ and precessing $\text{IMRPhenom}$) is $0.28_{-0.21}^{+0.35}$ and $0.35_{-0.27}^{+0.45}$, respectively \cite{abbott2016improved}. We can classify GW150914 as a GW event with moderate spins based on the mean value of $\chi_{p}$ and the wide range of 90\% credible bounds on the posteriors. This classification aligns with our classifier's outcome as a GW signal from a precessing event. Although our classifier does not provide further information on the scale of spins (low, medium, or high), it is sufficient from a search perspective to identify whether it is a precessing or aligned system. To comment on the precession level, one needs to perform parameter estimation via Bayesian inference to obtain the hard bounds on the posterior of $\chi_{\text{eff}}$ and $\chi_{\text{p}}$. As a future proposal, we could add another level of classification to determine the magnitude of precession - idetifying the precessing events as with low, medium and high precesssion. \\
\begin{table}[hbt]
\centering
\begin{tabular}{|c|c|c|c|c|c|}
\hline
Events &  $\text{M}(\text{M}_{\odot})$& \text{q} & $\text{L}1$ & $\text{H}1$ \\
\hline
$\texttt{GW}150914$ & 66.2& 1.16& $\BBH$ \, (\textcolor{blue}{$\PS$}) & $\BBH$ \, (\textcolor{blue}{$\PS$})  \\
$\texttt{GW}170104$ & 50.8 & 1.54 & $\BBH$ \, (\textcolor{blue}{$\AS$}) & $\BBH$ \, (\textcolor{blue}{$\AS$}) \\
$\texttt{GW}170608$& 18.6 & 1.44 & $\BBH$ \, (\textcolor{blue}{$\AS$}) & $\BBH$ \, (\textcolor{blue}{$\AS$}) \\
$\texttt{GW}170814$ & 55.8& 1.21& $\BBH$ \, (\textcolor{red}{$\AS$}) & $\BBH$ \, (\textcolor{red}{$\PS$}) \\
\hline
\end{tabular}
\caption{ 
The predicted $\BBH$ events with network-$\SNR \geq 13$ from the GWTC-1 are shown. The last two columns show the classified signals and their corresponding signal categories (AS: aligned spin, PS: precessing spin) using two $\CNN$ models; one is trained with $\text{L}1$ noise, and another is with $\text{H}1$ detectors noise. The blue(red) colors represent agreement(disagreement) between the CNN model's prediction.}
\label{tab:GWTC1}
\end{table}

\textbf{Events from GWTC-2:}
We tested nine GW events (See Table-\ref{tab:GWTC2}) from the second catalog, GWTC-2, with our trained NN model. Four events ( $\text{GW}190915\_{235702}$, $\text{GW}190519\_{153544}$, $\text{GW}190412$, $\text{GW}190521$), and two ($\text{GW}190728\_{064510}$, $\text{GW}190828\_{063405}$) have been identified as the precessing and aligned signals respectively. The prediction from both CNN models (trained with H1 and L1 noise) is consistent with each real event data from H1 and L1. However, for $\text{GW}190408\_{181802}$, $\text{GW}190707\_{093326}$, our trained classifiers show disagreement in their predictions. The events GW190521 \cite{estelles2022detailed} and GW190412 \cite{rodriguez2020gw190412} were reported to indicate the precession with $\chi_{p}$ posterior distribution constrained away from zero. Our prediction for these two events is consistent with several published works ~\cite{estelles2022detailed, rodriguez2020gw190412, precessionSNR, hoy2022evidence}. \\

\textbf{Events from GWTC-3}
We only analyzed four events from GWTC-3. Our classifiers identified all four events as GW signals from the precessing systems (See Table-\ref{tab:GWTC3}). $\text{GW}200129\_{065458}$ has been reported to have an inferred $\chi_{\text{p}}$ of $0.54^{+0.39}_{-0.39}$ \cite{gwtc3}. However, inference is shown to be sensitive to the choice of the waveform model. The $\chi_{\text{p}}$ posteriors are broad and uninformative for the other events \cite{gwtc3}. Thus, the precession effects of those events can not be disregarded. 

\begin{table}[hbt]
\centering
\begin{tabular}{|c|c|c|c|c|c|}
\hline
Events &  $\text{M}(\text{M}_{\odot})$ & \text{q} & $\text{L}1$ & $\text{H}1$ \\
\hline
$\text{GW}190408\_{181802}$ &  43.4& 1.34& $\BBH$ \, ({\color{red}{$\AS$}}) & $\BBH$ \, ({\color{red}{$\PS$}})  \\

$\text{GW}190521\_{074359}$ &  76.3& 1.29& $\BBH$ \, ({\color{blue}{$\PS$}}) & $\BBH$ \, ({\color{blue}{$\PS$}}) \\
$\text{GW}190707\_{093326}$ &  20.1 & 1.53& $\BBH$ \, ({\color{red}{$\PS$}}) & $\BBH$ \, ({\color{red}{$\AS$}}) \\
$\text{GW}190728\_{064510}$ &  20.7& 1.56& $\BBH$ \, ({\color{blue}{$\AS$}}) & $\BBH$ \, ({\color{blue}{$\AS$}}) \\
$\text{GW}190828\_{063405}$ &  57.2& 1.23& $\BBH$ \, ({\color{blue}{$\AS$}}) & $\BBH$ \, ({\color{blue}{$\AS$}}) \\
$\text{GW}190915\_{235702}$ &  57.2& 1.33& $\BBH$ \, ({\color{blue}{$\PS$}}) & $\BBH$ \, ({\color{blue}{$\PS$}}) \\
$\text{GW}190519\_{153544}$ & 105.6 & 1.59& $\BBH$ \, ({\color{blue}{$\PS$}}) & $\BBH$ \, ({\color{blue}{$\PS$}}) \\
$\text{GW}190412$ & 36.8 &3.07& $\BBH$ \, ({\color{blue}{$\text{PS}$}}) & $\BBH$ \, ({\color{blue}{$\text{PS}$}}) \\
$\text{GW}190521$ &  153.1 & 1.72& $\BBH$ \, ({\color{blue}{$\text{PS}$}}) & $\BBH$ \, ({\color{blue}{$\text{PS}$}}) \\
\hline
\end{tabular}
\caption{The predicted BBH events with network $\SNR \geq 13$ from the GWTC-2 for both L1 and H1 detectors are shown. Out of nine tested events, the prediction of CNN models is consistent for seven events.}
\label{tab:GWTC2}
\end{table}
\begin{table}[hbt]
\centering
\begin{tabular}{|c|c|c|c|c|c|}
\hline
Events &  $\text{M}(\text{M}_{\odot})$ & \text{q} & $\text{L}1$ & $\text{H}1$ \\
\hline
$\text{GW}191109\_{010717}$ &  112& 1.38 & $\BBH$ \, ({\color{blue}{$\PS$}}) & $\BBH$ \, ({\color{blue}{$\PS$}}) \\
$\text{GW}200129\_{065458}$ &  63.9& 1.25 & $\BBH$ \, ({\color{blue}{$\PS$}}) & $\BBH$ \, ({\color{blue}{$\PS$}}) \\
$\text{GW}200224\_{222234}$ &  72.3 & 1.22& $\BBH$ \, ({\color{blue}{$\PS$}}) & $\BBH$ \, ({\color{blue}{$\PS$}}) \\
$\text{GW}200311\_{115853}$ &  61.9& 1.23& $\BBH$ \, ({\color{blue}{$\PS$}}) & $\BBH$ \, ({\color{blue}{$\PS$}}) \\
\hline
\end{tabular}
\caption{A specific set of $\BBH$ events with network $\SNR \geq 13$ from GWTC-3 is used to test the performance of our trained Neural Network model. The prediction of two CNNs (one trained with H1 and another with L1 noise) is identical for the chosen GW events.}
\label{tab:GWTC3}
\end{table}
\section{Discussion \& Conclusion}
\label{Sec:Conclusion}
\textbf{Outlook:} This work has explored the possibility of detecting GW signals from precessing BBH systems using convolutional neural networks. We observed that the CNN model can effectively detect GW signals from precessing spin against pure noise. However, distinguishing between aligned and precessing signals is challenging due to their similar morphologies, which can confuse the Neural Network and increase the misclassification rate. 

Our study found that classification accuracy improved when the CNN model was trained with varied signal morphologies by analyzing the signal parameter space. Specifically, adjusting the $\theta_{\text{JN}}$ and $\phi_{\text{JL}}$ values helped us understand how signals differ in morphology. A small $\theta_{\text{JN}}$ value led to a high misclassification rate because the morphologies of aligned and precessing signals were very similar. Conversely, higher $\theta_{\text{JN}}$ values improved classification accuracy.

We tested our CNN model against synthetic data and publicly available real events from the first three observational runs from both the Hanford and Livingston detectors. The trained CNN model achieved an overall accuracy of more than $97\%$ in the case of simulated noise and $\sim 95\%$ in the case of real noise for distinguishing precessing signals from pure noise and aligned signals. We employed a hierarchical strategy with two identical CNN models: the first model distinguished between pure noise and noisy BBH signal, while the second model further classified the signals detected from the first stage as aligned or precessing signals. Testing on real BBH events (with network SNR $\geq 13$) from the GW catalogs showed that all events were correctly classified as signals by the first CNN model, with the second model's classifications generally consistent with reported events in LVK catalogs and other research. \\
\textbf{Limitations and Future Scope:} 
We observed that the individual predictions from the second CNN model for the two detectors (H1 and L1) differed for some real events. Since the CNN models for each detector were trained independently, differing predictions for the same event are expected. Given that the L1 detector is currently more sensitive, we rely on its results in case of disagreement between H1 and L1. However, we are investigating this issue further to develop a more robust solution.

Additionally, we used identical Neural Network models for all case studies, regardless of the noise model and classification stage. This approach leaves room for developing new, specialized models tailored to different case studies. While this work primarily aimed to assess the CNN model's sensitivity in distinguishing between aligned and precessing systems, designing advanced models for these specific scenarios is beyond the scope of this study and is suggested for future research.

We also used a residual neural network (ResNet) for these case studies, achieving performance similar to that of the CNN. This work is ongoing, and extending the training dataset may be necessary to improve ResNet's accuracy. In future studies, we will provide a detailed comparison between ResNet and CNN for the cases presented in this paper.

Furthermore, this work explored obtaining coincident triggers from a multi-detector study. We conducted a multi-detector coincidence test to identify the coalescence time of signals within data chunks. Once the coalescence time of a signal is identified, the classifier can determine whether the corresponding signal is aligned or precessing. \\
\textbf{Advatanges:} 
Identifying GW signals using a Neural Network offers several advantages:
\begin{itemize}
\item High Accuracy for Precessing Signals: The CNN model can detect highly precessing GW signals with high accuracy, which may be missed by aligned template-based matched filter search methods. Classical search pipelines, which currently use aligned spin template banks, can detect precessing signals only if their morphologies are similar to those of aligned signals (as the match between aligned and precessing waveforms will be better). In contrast, our strategy is inherently more effective at distinguishing between precessing and aligned signals when morphologies differ significantly (when the match between aligned and precessing waveforms is poor).
\item Detecting highly precessing signals typically requires a larger precessing template bank than an aligned bank, resulting in increased computational costs for matched-filter-based searches. Furthermore, efficient template placement algorithms for precessing signals are not well-established. CNN-based methods, which do not rely on a template bank, can significantly reduce the computational cost of searches for precessing spin systems.
\item Real-Time Detection Potential: This work represents the first step towards developing a Neural Network-based framework for real-time detection of precessing BBH systems. The approach is easily extendable to other binary systems, such as neutron star-black hole (NSBH) systems. Future work will incorporate signal representations from other compact binary sources to enhance the model and thoroughly analyze detected GW events from the last three observational runs. 
\end{itemize}

\vspace{0.5cm}

\begin{acknowledgments}
C.V is thankful to the Council for Scientific and Industrial Research (CSIR), India, for providing the senior Research Fellowship. A.R is supported by the research program of the Netherlands Organisation for Scientific Research (NWO). S.C is supported by the National Science Foundation under Grant No. PHY-2309332. They are grateful for computational resources provided by the LIGO Laboratory and supported by the National Science Foundation Grants No. PHY-0757058 and No. PHY-0823459. This material is based upon work supported by NSF's LIGO Laboratory which is a major facility fully funded by the National Science Foundation. 
\end{acknowledgments}
%
\section*{References}
\bibliography{references}
\section{Appendix}
\begin{figure}[!ht]
\centering
\includegraphics[scale = 0.3]{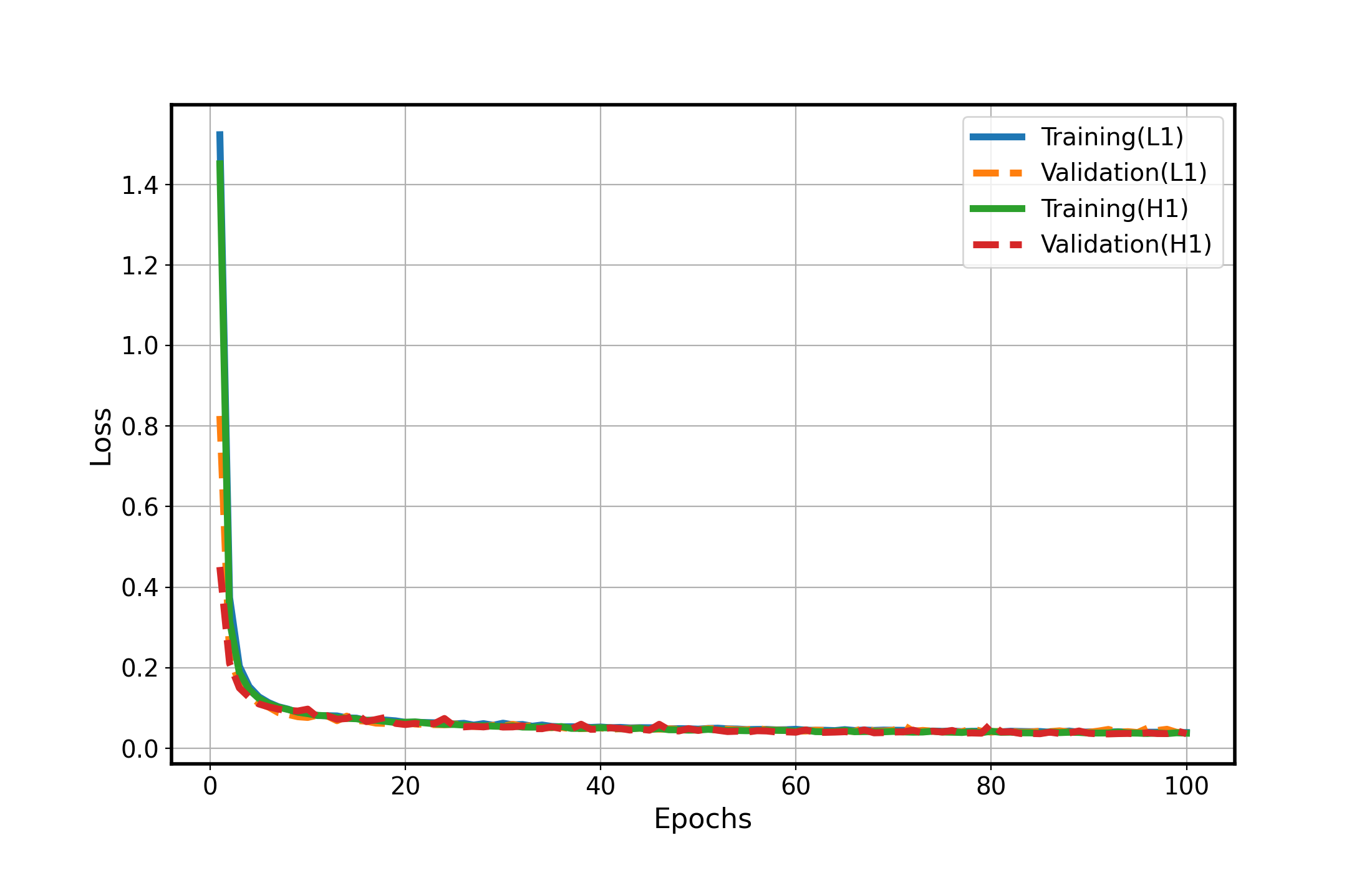} %
\qquad
\includegraphics[scale = 0.3]{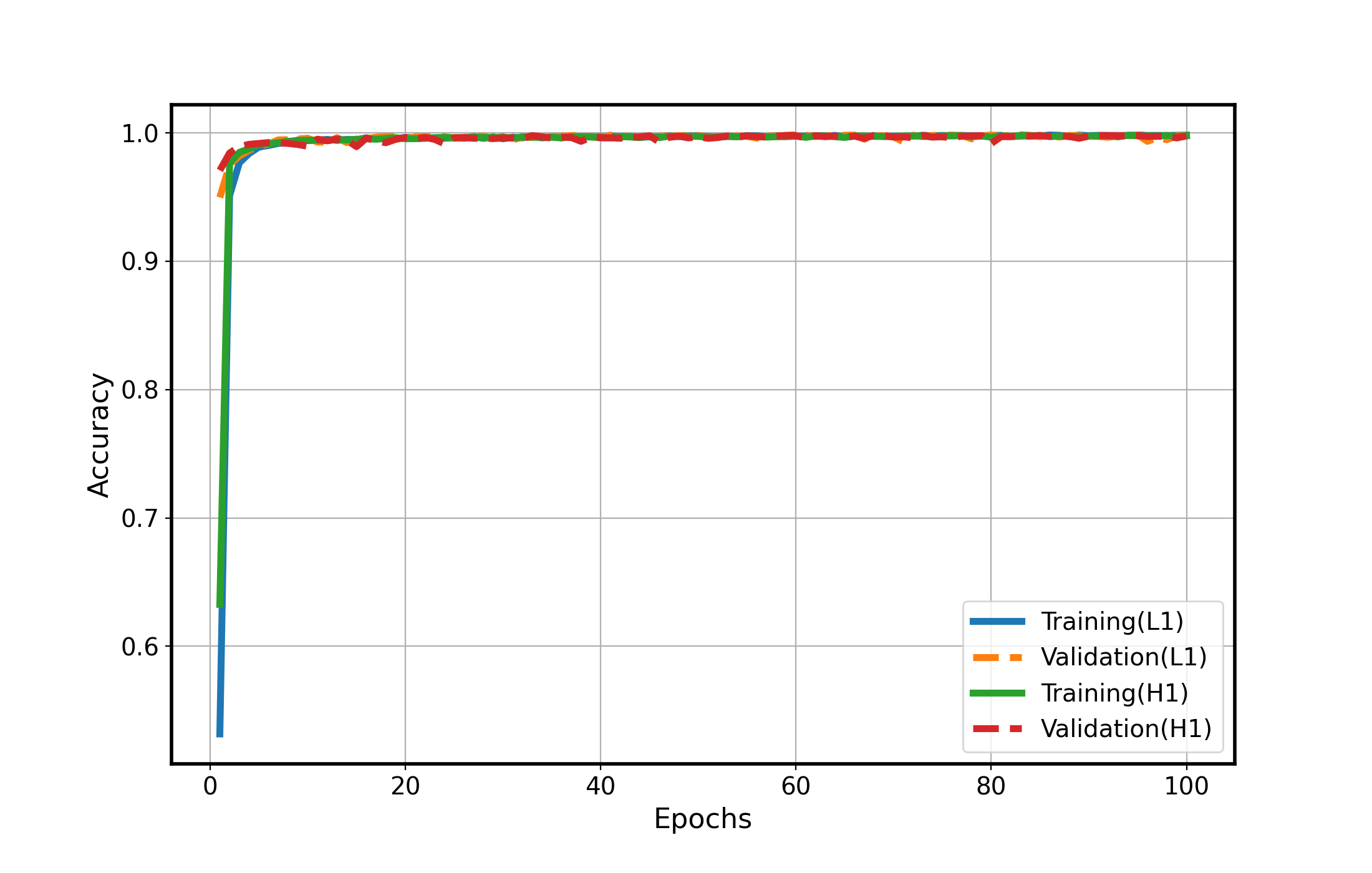} %
\caption{Figures (a) and (b) show the training and validation loss and accuracy with the number of epochs of the classifier for noise and signal classification for detector noise. }%
\label{Fig:acc_loss_plot}%
\end{figure}
\begin{figure}[!ht]
\centering
\includegraphics[scale=0.3]{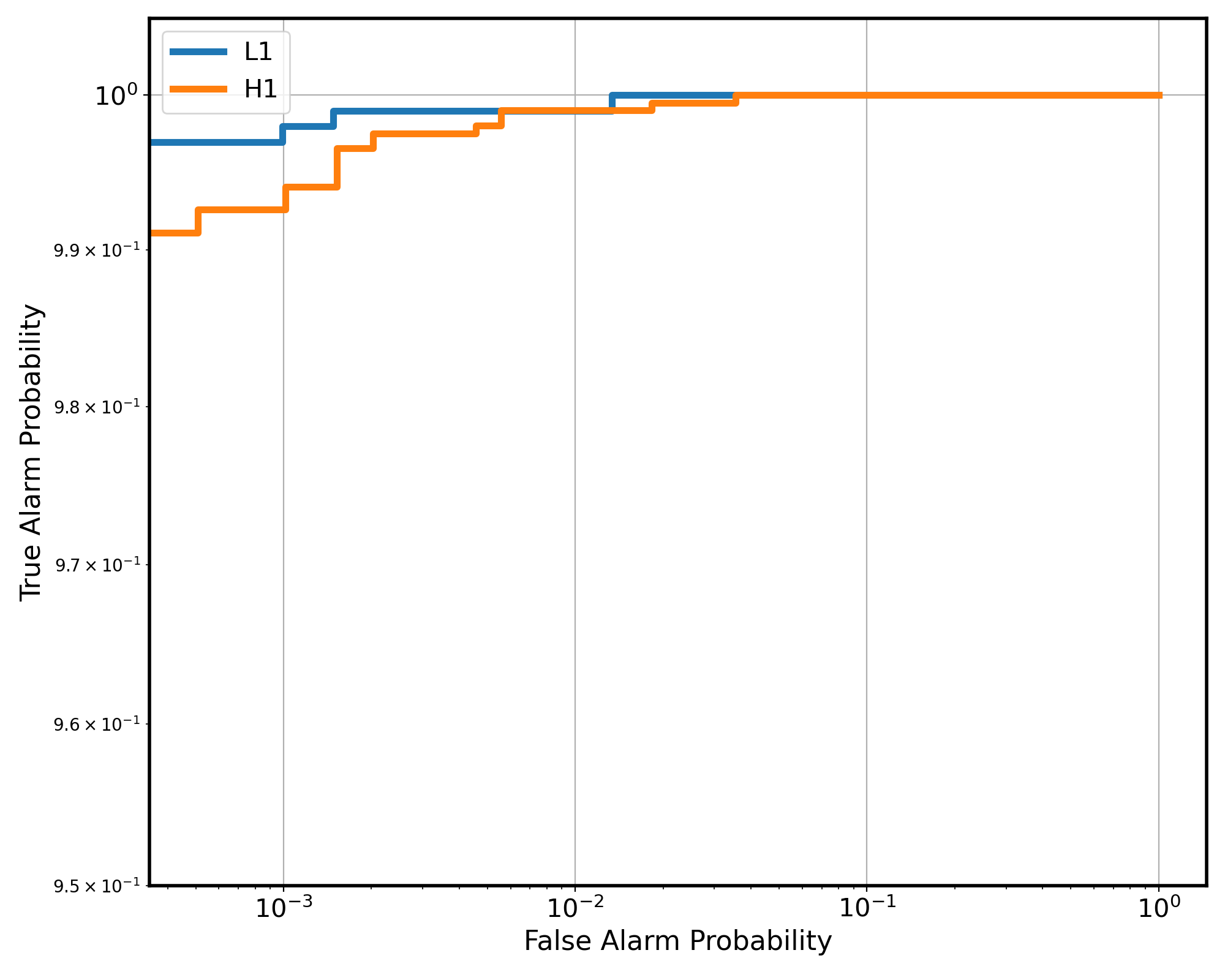}%
\caption{The figure shows the Receiver Operating Characteristic (ROC) of the classifier for classifying noise and signal for the detector noise}
\label{Fig:roc_signal}%
\end{figure}
\subsection*{Classification Accuracy Over Parameter Space}
\begin{figure}[!ht]
\centering
\includegraphics[width=0.35\textwidth]{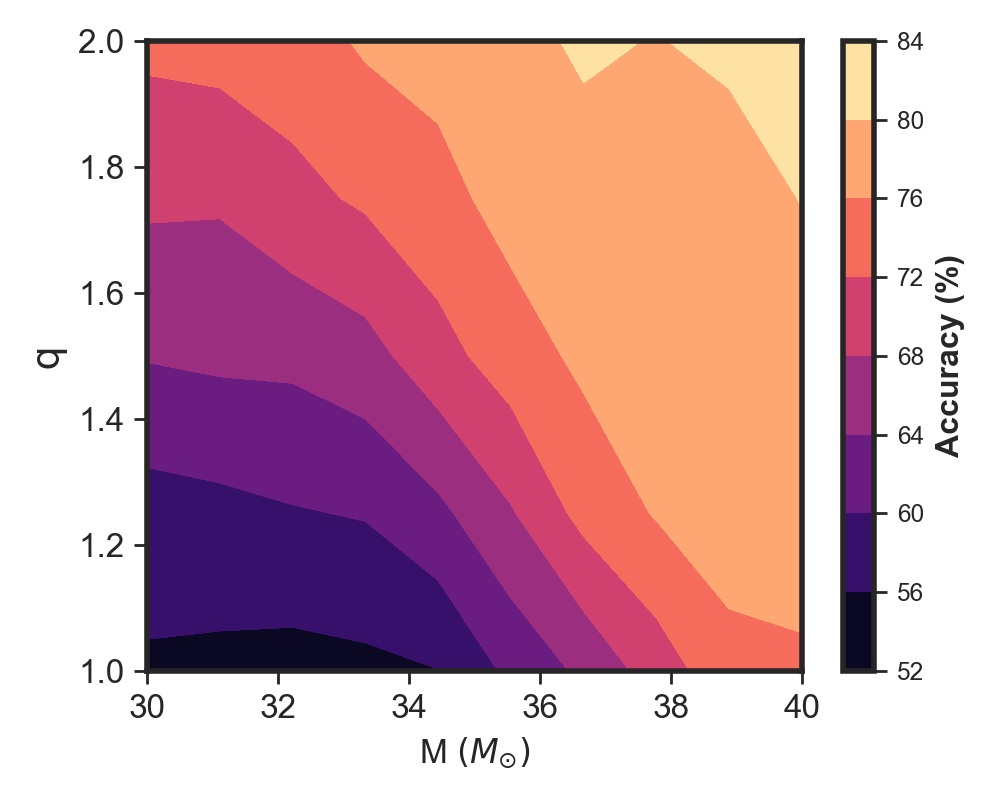} %
\qquad
\includegraphics[scale = 0.5]{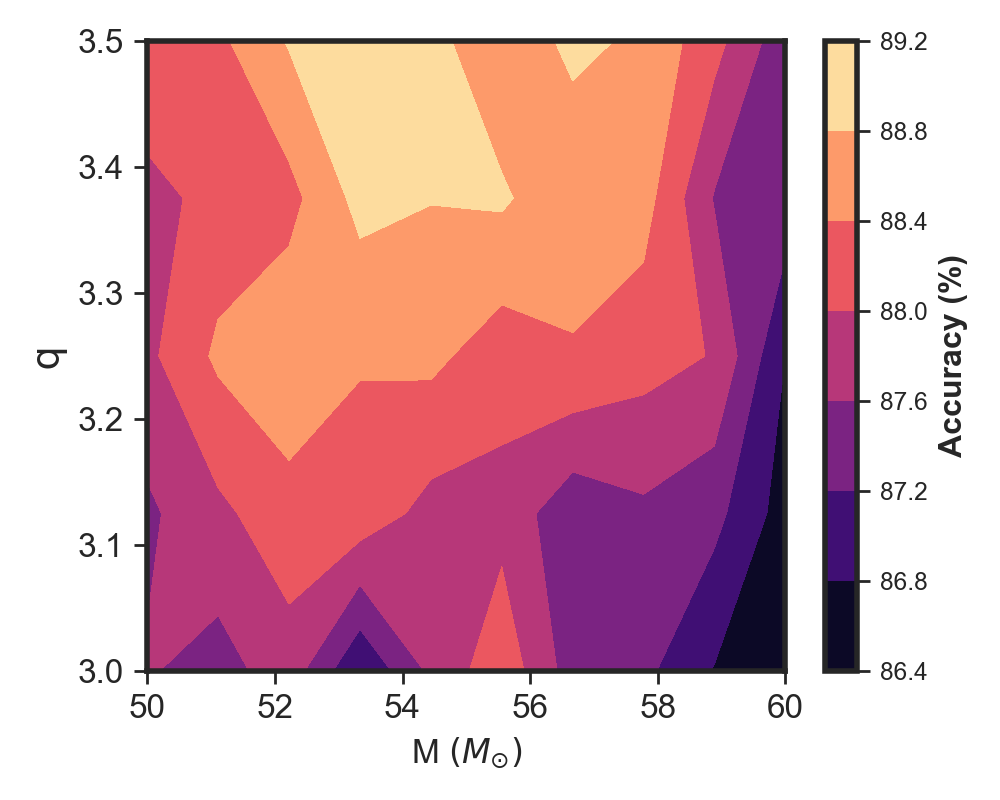} 
\caption{Sub-figures $(a) \& (b)$ show that the average accuracy of signal classification (aligned vs. precessing) varies in the different grids on total mass and mass ratio. }%
\label{Fig:Acc_M_q}%
\end{figure}

\begin{figure}[!ht]
\centering
\includegraphics[scale = 0.5]{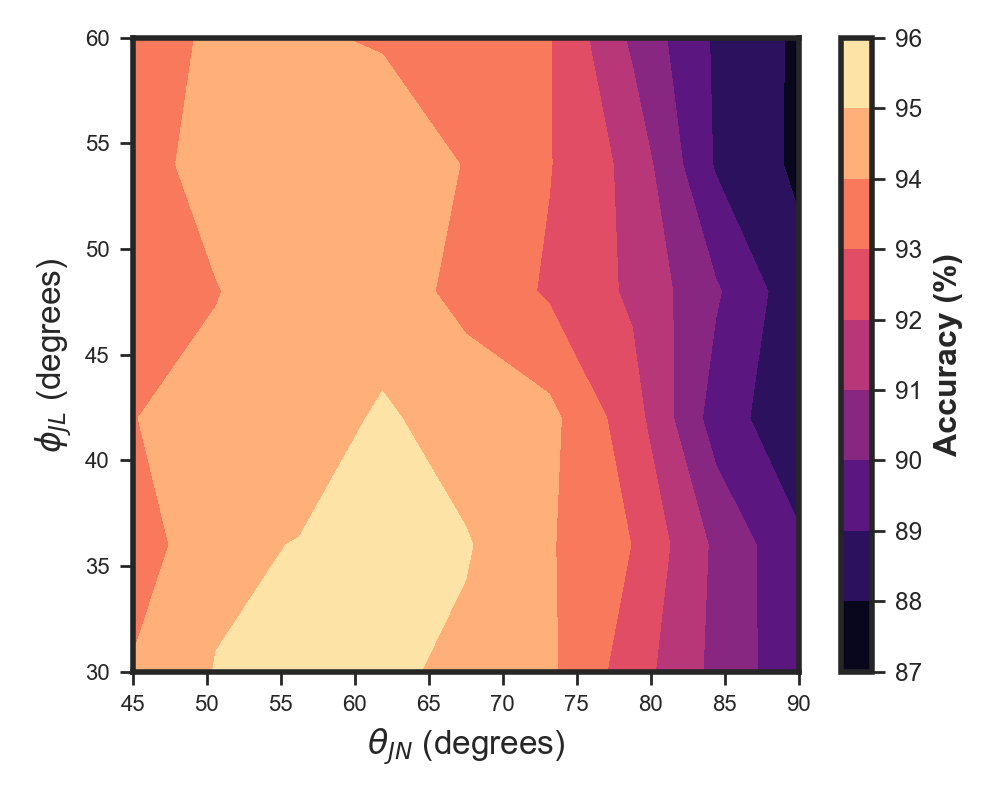} %
\qquad
\includegraphics[scale = 0.5]{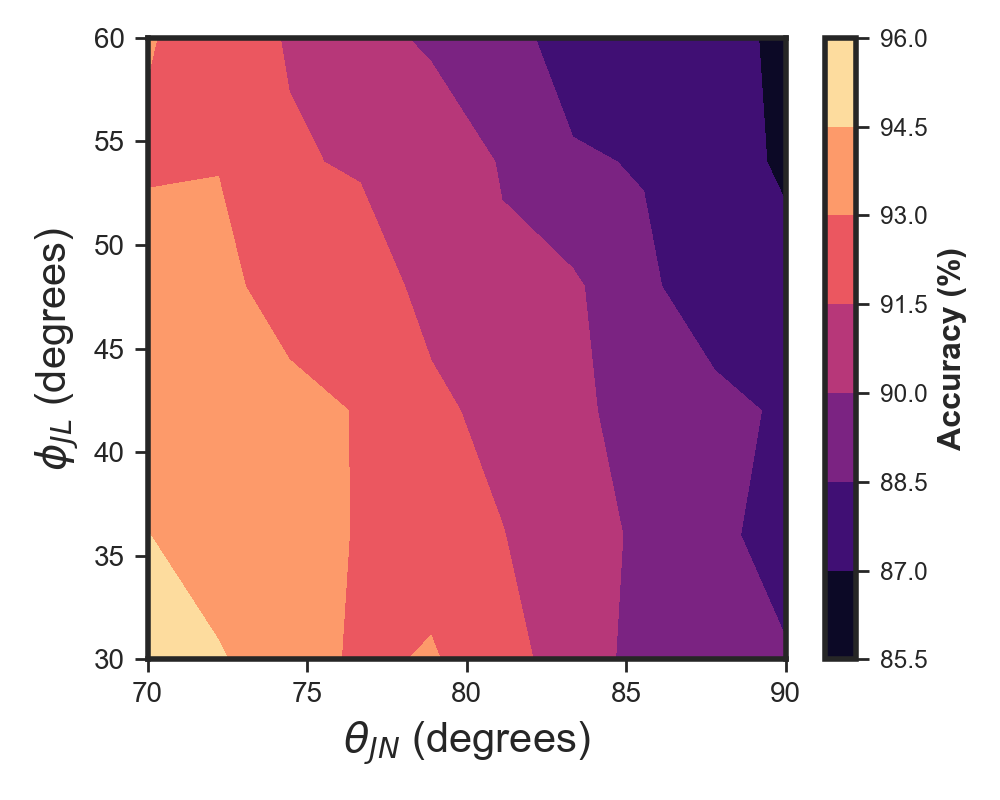} %
\caption{Accuracy of signal classification in terms of aligned vs. precessing varies in the different grids on $\theta_{\text{JN}}$ and $\phi_{\text{JL}}$.}%
\label{Fig:Acc_theta_phi}%
\end{figure}

The classifier performs well at the first stage (Fig \ref{Fig:acc_loss_plot}, Fig \ref{Fig:roc_signal}) as it is independent of the parameter space but it is not the same case for the classifier at the second stage. Hence, we explored the accuracy of our classifier at stage second for different parts of the parameter spaces. In general, the total number of parameters is sixteen (including intrinsic and extrinsic parameters), and training and testing signals are generated by varying all the parameters. However, the classifiers' accuracy can vary based on aligned and low (high) precessing signals. Therefore, we vary total mass, mass ratio, $\theta_{\text{JN}}$ and $\Phi_{\text{JL}}$ to generate low (high) precessing signals and test the accuracy of the classifier. Figure \ref{Fig:Acc_M_q} shows the accuracy for two different total mass and mass ratio ranges. One in between $M \in [30-40] M_{\odot}$ and $q \in [1-2]$ and another in between $ M \in [50-60] M_{\odot}$ and $q \in [1-3.5]$. The accuracy is high with high total mass and mass ratio. Similarly Figure \ref{Fig:Acc_theta_phi} shows accuracy for fixed range of $\theta_{\text{JN}} \in [\frac{\pi}{4}, \frac{\pi}{2}]$ and $\phi_{\text{JL}} \in [\frac{\pi}{6}, \frac{\pi}{3}]$. The sub-figure (right side) shows the accuracy for specific range for $\theta_{\text{JN}} \in [\frac{7\pi}{18} , \frac{\pi}{2}]$ and $\phi_{\text{JL}} \in [\frac{\pi}{6}, \frac{\pi}{3}]$. From, all the test examples, we conclude that the accuracy of our classifier is high when the test signals are generated by varying $\theta_{\text{JN}} \in [\frac{\pi}{4}, \frac{5\pi}{12}]$ and $\phi_{\text{JL}} \in [0, \frac{\pi}{3}]$. 
\subsection*{Multi-Detector Coincident Framework}
\begin{figure}[!ht]
\centering
\includegraphics[width=0.5\textwidth]{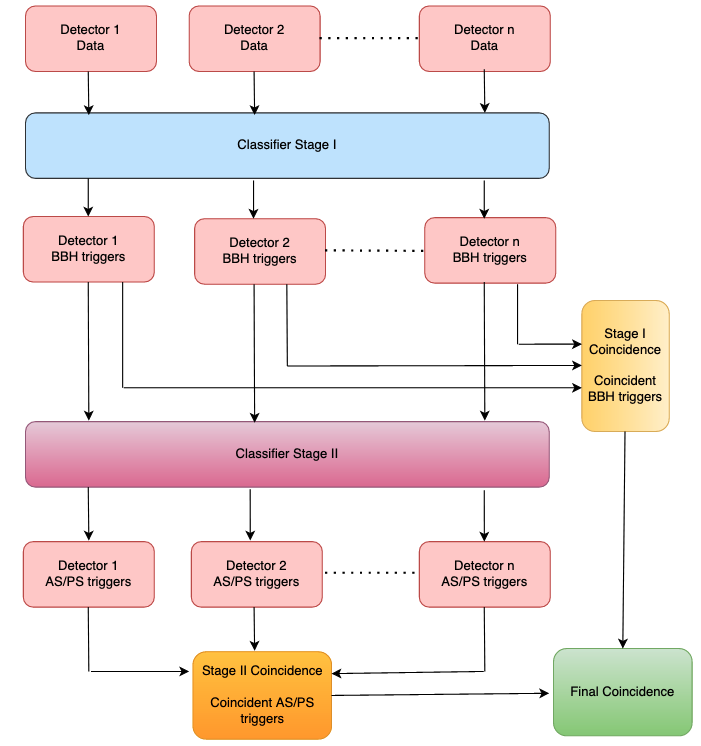}
\caption{Schematic diagram of the coincident test framework using CNN classifiers for the multi-detector scenario. A continuous data stream obtained from individual detectors passed through the pre-trained classifiers in the first stage to identify the possible data chunks of interest. The data chunks are identified by thresholding the detection probability (softmax value). These data chunks from each detector then go through a coincidence test to obtain the coalescence time. Further, the data chunks from individual detectors are passed to the second pre-trained classifier for classification into AS or PS, named stage II. If all the classifiers identify the event as AS or PS, then only we call them coincident AS or PS triggers. 
In some cases, the prediction may differ due to the independent nature of the training of Neural Network models. We found this issue is persistent while testing for real GW events (e.g., GW170814, GW190408, GW190707) in Table-\ref{tab:GWTC1} and \ref{tab:GWTC2}. Then, we can not conclude whether the detected event is coming from AS or PS systems. It is a limitation of our approach. However, the first stage coincident test still provides the correct estimation of the coalescence time and specific data chunk with signal presence. }
\label{fig:coincident-framework}
\end{figure}

In our proposed framework, we need to perform two independent coincident tests between our trained Neural Network models (a total of four independent models) for individual detectors, one for obtaining the coalescence time for the vital data chunks to detect the trace of GW signal and the second for receiving the confirmation on signal categories as an aligned or precessing one. We would first perform the coincident test related to signal against noise prediction, which will help to obtain the coalescence time. The second coincident test to confirm the nature of the signal category would work perfectly when we need to predict between aligned and highly precessing signals. However, the coincident tests may fail to recognize aligned and low-precessing signals. The proposed framework is shown in Figure \ref{fig:coincident-framework}.
\subsection*{Coincident Test on Real GW Events}
\begin{figure*}[!ht]
\centering
\includegraphics[width=0.75\textwidth]{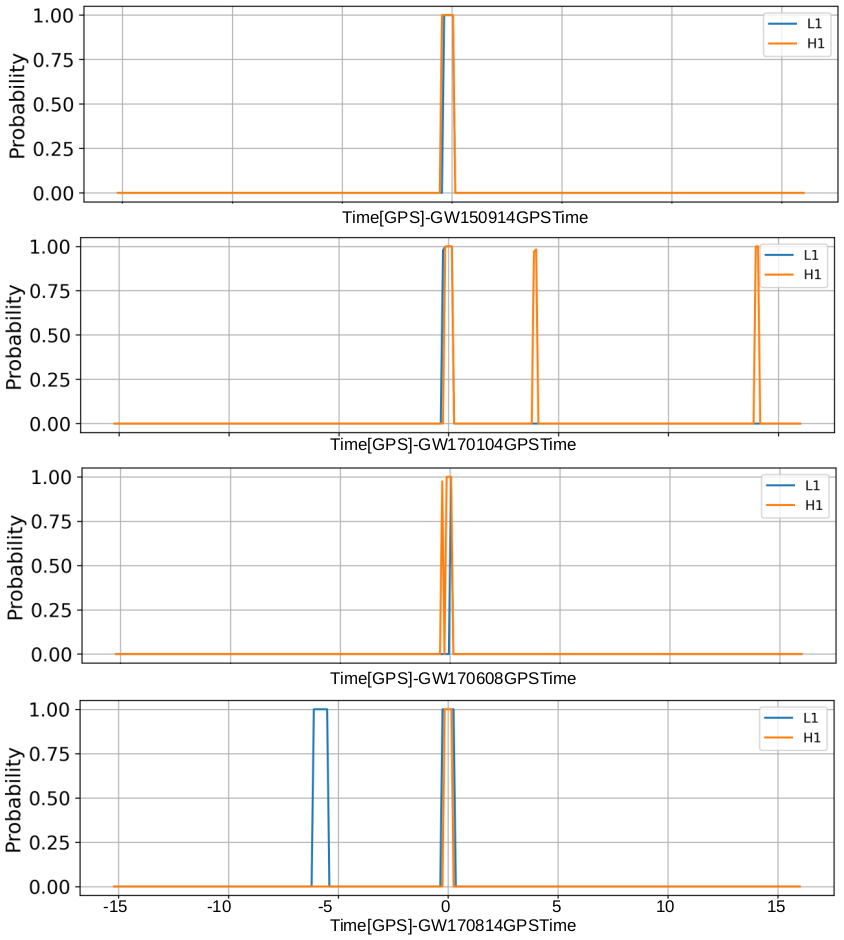}
\caption{The coincident test for a few GW BBH events from GWTC-1 is shown. The events are chosen based on their network $\SNR \geq 13.0$.}
\label{fig:ProbGWTC1}
\end{figure*}

\begin{figure*}[!ht]
\centering
\includegraphics[width=0.7\textwidth]{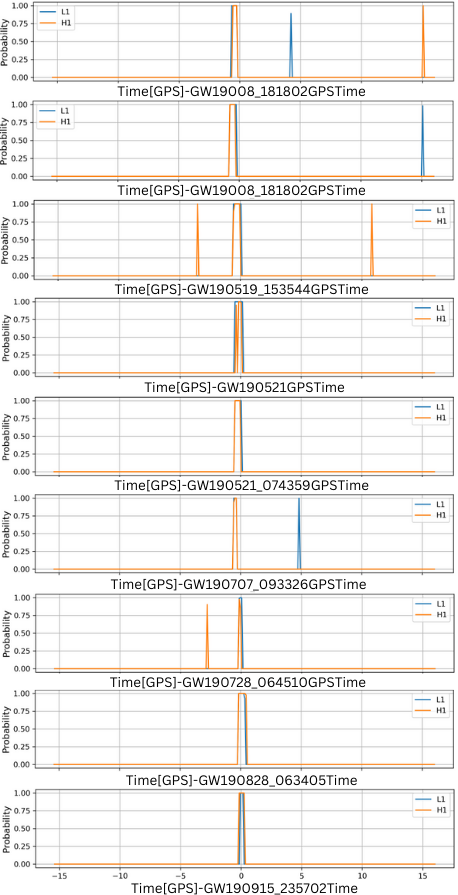}
\caption{A specific set of events with network $\SNR \geq 13.0$ from GWTC-2 is used to demonstrate the coincident test scheme.}
\label{fig:ProbGWTC2}
\end{figure*}

\begin{figure*}[!ht]
\centering
\includegraphics[width=0.75\textwidth]{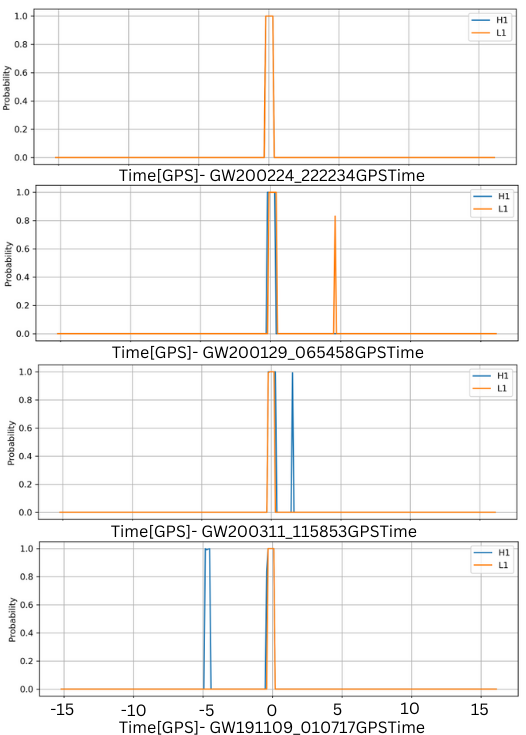}
\caption{The coincident test results for BBH events from GWTC-3 with network $\SNR \geq 13.0$ are shown.}
\label{fig:ProbGWTC3}
\end{figure*}

We have performed a coincident test using our trained Neural Network models on all the real events shown in Tables-\ref{tab:GWTC1}, \ref{tab:GWTC2}, \ref{tab:GWTC3}. The coincident test for those events are shown in Figures \ref{fig:ProbGWTC1}, \ref{fig:ProbGWTC2}, \ref{fig:ProbGWTC3}. 
\end{document}